\documentclass[%
aps,%
prb,
twoside,
twocolumn,
showpacs,%
preprintnumbers,
floatfix%
]{revtex4}

\usepackage{dcolumn}
\usepackage{bm}
\usepackage{bbm}
\usepackage{ifthen}
\usepackage[usenames]{color}
\usepackage{amssymb}
\usepackage{amsfonts}
\usepackage{amsmath}
\usepackage{graphicx}
\usepackage[colorlinks,urlcolor=green,citecolor=blue,pagecolor=violet]{hyperref}

\def\ee{{\rm e}}
\def\ii{\textrm{i}\,\!}

\newcommand{\lsb} {\left[}
\newcommand{\rsb} {\right]}
\newcommand{\lrb} {\left(}
\newcommand{\rrb} {\right)}
\newcommand{\lcb} {\left\{}
\newcommand{\rcb} {\right\}}
\newcommand{\lab} {\left\langle}
\newcommand{\rab} {\right\rangle}
\newcommand{\lnothing} {\left.}
\newcommand{\rnothing} {\right.}
\newcommand{\ImP} {\textrm {Im}}
\newcommand{\vect} {\mathbf}
\newcommand{\vectGr} {\boldsymbol}
\newcommand{\abso}[1]{\left\vert #1 \right\vert}
\newcommand{\ket}[1]{\vert #1 \rangle}
\newcommand{\bra}[1]{\langle #1 \vert}

\newcommand{\rket}[1]{\vert #1\, )}
\newcommand{\rbra}[1]{(\, #1 \vert}

\newcommand{\Ham} {\ensuremath{\mathcal H}}

\newcommand{\CNT}{CNT }
\newcommand{\CNTns}{CNT}
\newcommand{\CNTs}{CNTs }
\newcommand{\CNTsns}{CNTs}

\newcommand{\SOI}{SOI }
\newcommand{\SOIns}{SOI}

\newcommand{\ie}{\textit{i.e.~}}
\newcommand{\eg}{\textit{e.g.~}}
\newcommand{\cf}{\textit{cf.~}}
\newcommand{\R}[1]{\mathrm{ #1}}
\newcommand{\eV} {e\R{V}}

\newcommand{\azim} {\theta}
\newcommand{\radio} {R}
\newcommand{\aC}{a_{C}}
\newcommand{\aone}{\vect{a}_1}
\newcommand{\atwo}{\vect{a}_2}

\newcommand{\bone}{\vect{b}_1}
\newcommand{\btwo}{\vect{b}_2}

\newcommand{\Kone}{\vect{K}}
\newcommand{\Ktwo}{\vect{K'}}
\newcommand{\chiralvec}{\vect{C}}
\newcommand{\translvec}{\vect{T}}

\begin{document}
\date{\today}

\title{Signatures of spin-orbit interaction in transport properties of finite carbon nanotubes in a parallel magnetic field}

\author{Miriam del Valle, Magdalena Marga\'{n}ska, Milena Grifoni}
\affiliation{Institute for Theoretical Physics, University of Regensburg, D-93040 Regensburg, Germany}

\begin{abstract}
The transport properties of finite nanotubes placed in a magnetic field parallel to their axes are investigated. Upon including spin-orbit coupling and curvature effects, two main phenomena are analyzed which crucially depend on the tube's chirality: i) Finite carbon nanotubes in a parallel magnetic field may present a suppression of current due to the localization at the edges of otherwise conducting states.
This phenomenon occurs due to the magnetic-field dependent open boundary conditions obeyed by the carbon nanotube's wave functions.
The transport is fully suppressed above threshold values of the magnetic field which depend on the nanotube chirality, length and on the spin-orbit coupling.
 ii) Reversible spin polarized currents can be obtained upon tuning the magnetic field, exploiting the curvature-induced spin-orbit splitting.
\end{abstract}


\maketitle
\section{Introduction}

Carbon nanotubes (\CNTsns) have been the focus of intense study in the last decades as they exhibit remarkable properties that make them good candidates for molecular electronic devices.~\cite{AvourisCP07,ReichTM04} Lately their application as building blocks in spintronics has also been addressed.~\cite{TsukagoshiAA99,KimSKK02,SahooKFHGCS05,CottetKSMCBBMS06,KollerMG07,SchenkeKMG09} Nonetheless the importance of spin-orbit coupling in these systems has been for a long time underestimated. The spin-orbit interaction (\SOIns) was believed to be very weak, due to the low atomic number $Z=6$ of carbon. As a result spin degenerate states were assumed. Only recently it was proven experimentally~\cite{KuemmethIRM08} that in the spectrum of ultraclean \CNTs the effects of this coupling between spin and orbital degrees of freedom are clearly visible. This observation is in agreement with previous theoretical predictions~\cite{Ando00,ChicoLM04,HuertasHernandoGB06} which argued that \SOI could be significant in \CNTs owing to their curvature. Understanding the effects of this coupling on the spectrum of \CNTs is essential for the successful manipulation of the different degrees of freedom of these systems.
Signatures of spin-orbit coupling in the magnetic field dependence of the spectrum have been further identified in measurements~\cite{ChurchillKHBRFSTWM09} of relaxation and dephasing times in a \CNT quantum dot and more recently in magnetoconductance experiments~\cite{JhangMSPWGvZWS10} performed on open \CNTns-wires, showing a double-peak feature caused by a spin splitting of the conduction bands due to the \SOIns. The observed spin-orbit energy splitting is compatible with previous experimental and theoretical values~\cite{KuemmethIRM08,Ando00,HuertasHernandoGB06,BulaevTL08} and further supports the relevance of spin-orbit effects in these systems.
Moreover new transport measurements~\cite{JespersenGPMFNF11} in \CNT quantum dots with magnetic fields both parallel and perpendicular to the nanotube serve to analyze the influence of \SOI also in the cotunneling regime.

The application of a parallel magnetic field modulates the distribution of the electronic spectrum of nanotubes,~\cite{AjikiA93,TianD94} as has
been confirmed experimentally in several optical and transport measurements.~\cite{BachtoldSSBFNS99,FujiwaraTSYU99,ZaricOKSMSHSW04,CoskunWVGB04,StrunkSR06,StojetzRMTFS07,FedorovBSJR10} Further theoretical studies~\cite{ZhangLD98,KrompiewskiGC04,NemecC06,KrompiewskiC07} reveal that the transport properties are directly related to the magnetic-field modulated density of states.
The \SOI-modified band structure and its implications on the electronic properties of \CNTs in a magnetic field has been in the focus of extensive theoretical work~\cite{BulaevTL08,IzumidaSS09,Wunsch09,SecchiR09,ZhouLD09,JeongL09,FangZL08,LoganG09,WeissRKCF10}.
However, at present only few theoretical works exist~\cite{JespersenGPMFNF11,MarganskadVJSG11} which investigate quantum transport through \CNTs including magnetic field, curvature and \SOI effects.

In this paper we analyze the transport properties of finite-size carbon nanotubes in a parallel magnetic field in the regime of strong nanotube-lead coupling, using a tight-binding Hamiltonian for the nanotube system and Green function techniques.~\cite{CunibertiFR05} A magnetic field parallel to the tube axis is included via the Aharonov-Bohm phase and Zeeman effect, and we take into account the spin-orbit coupling, which is responsible for the appearance of a spin-splitting  and of spin-flip processes. The spin-splitting may allow for preparation of states with controlled spin and valley (the so-called isospin) degrees of freedom.  The inclusion of the \SOI and the magnetic field leads furthermore to the breaking of the electron-hole symmetry, as we notice in our analysis of these systems. Without an electron-hole symmetry spin-up and spin-down states may cross the Fermi energy at different magnetic fluxes giving rise to peaks in the magnetoconductance corresponding to spin-polarized currents around the Fermi energy similar to those experimentally observed in Ref.~\onlinecite{JhangMSPWGvZWS10} and as identified in the present work.

For zigzag and chiral nanotubes the finite size of the systems leads to a localization of the states in the magnetic field with a consequent suppression of current.~\cite{SasakiMSK05,MarganskadVJSG11} We show that, as a consequence of the \SOI, this localization starts at a threshold value of the magnetic field which is different for the two spin species. This value can be further tuned with the chirality, radius or length of the tube, allowing for controlled polarized states. Furthermore we find that numerical transport calculations in finite-size \CNTs are to a very good extent reproduced by using an analytical model Hamiltonian in the reciprocal space. The proper boundary conditions are crucial for the understanding of finite-size effects. The analytical approach provides as well a fundamental insight into the nature of the states taking part in the transport processes.

Part of the discussion on localization was presented in abridged form in Ref.~\onlinecite{MarganskadVJSG11}. The scope of the present work is to analyze various consequences of \SOI on the transport characteristics of \CNTs strongly coupled to leads.

The paper is organized as follows. In Sec.~\ref{sec_theory} we introduce the tight-binding approach that we use for modelling our system, where we keep the interactions to nearest neighbors only because, despite its simplicity, this approximation is known to provide a good agreement with experiments.~\cite{ReichTM04} Our original tight-binding Hamiltonian is modified to consecutively include the effects of an applied axial magnetic field, the curvature of the nanotube and the spin-orbit coupling.
Starting from this generalized tight-binding model for the real-space description of the \CNTsns, an effective Hamiltonian for the $\pi$ bands is obtained, which is formally the same as the one derived previously by other groups.~\cite{Ando00,HuertasHernandoGB06} This Hamiltonian will allow us in Sec.~\ref{sec_model:dispersion_BC} to make predictions for the energy spectrum when taking into account the boundary conditions to be satisfied at the \CNT ends. Finally the transport calculations are shown and discussed in Sec.~\ref{sec_numerical_transp}-\ref{sec_numerical_pol}. An analysis of the contribution to the \CNT conductance properties of the different effects under consideration is carried out in Sec.~\ref{sec_numerical_transp}, followed by a comparison of numerical and analytical results and an examination of the localization in the magnetic field of previously extended states in Sec.~\ref{sec_numerical_loc}. The great potential of these \CNT systems for their use in spintronics is demonstrated in Sec.~\ref{sec_numerical_pol}, where the spin-polarized currents at small bias voltages are investigated.
A summary of the main results is found in Sec.~\ref{sec_concl}.

\section{Magnetic field and spin-orbit coupling effects}\label{sec_theory}

\begin{figure}[h!]
\centerline{\includegraphics[width=\linewidth]{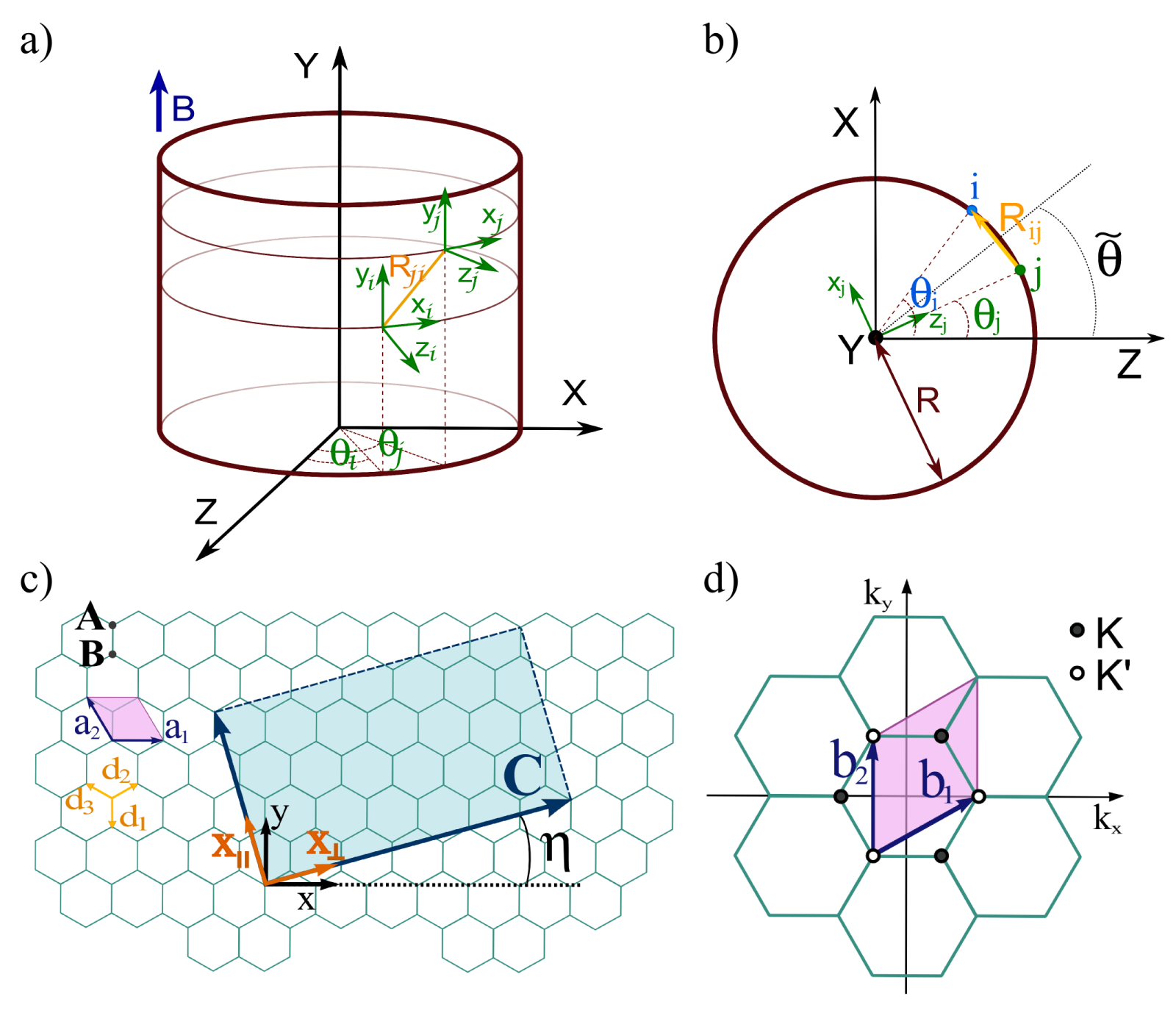}}
\caption{\label{fig:coord}
System and coordinates. a) Schematic illustration of the carbon nanotube with the local and global coordinate systems used for its description. The direction of the magnetic field is parallel to the tube axis.
b) Projection of the nanotube onto the $X-Z$ plane. $i$ and $j$ schematically show the projection of two nearest neighbor atoms (at an exaggerated distance in order to visualize the different variables). The angles are measured from the positive $Z$ axis and $\abso{\theta_j-\theta_i}$ is very small for large radii, so that in fact both angles can be seen as $\tilde{\theta}$. $\radio$ is the radius of the nanotube, whereas $\vect{R}_{ij}$ is the vector connecting the two neighboring atoms (here we plot actually the projection of this vector).
c) Real space representation of a fragment of a graphene lattice with translational vectors $\aone$ and $\atwo$. The chiral vector is $\chiralvec$ and the unit vectors $\hat{x}_\perp$ and $\hat{x}_\parallel$ reflect the parallel and perpendicular directions with respect to the nanotube axis. The three $B$ nearest neighbor atoms of an atom $A$ are located at positions given by $\vect{d}_l$ with $l=1,\ 2,\ 3$.  d) Reciprocal space for the graphene, spanned by the vectors $\bone$ and $\btwo$. The central hexagon is the symmetric Brillouin zone. Its corners are the so-called $\Kone$ and $\Ktwo$ points or Dirac points.}
\end{figure}

We wish to describe a finite single-wall carbon nanotube with a chiral vector~\cite{notation} $\chiralvec$ as sketched in Fig.~\ref{fig:coord}. The chiral vector uniquely defines the nanotube and determines its unit cell, radius $\radio$ and other properties such as its chiral angle $\eta$, which is defined such that for zigzag \CNTs $\eta=0$.

Neglecting for the moment curvature effects and spin-orbit interaction, we restrict ourselves to one $2p$ orbital ($p_z$) per site, as these orbitals give rise to the $\pi$ molecular orbitals, which in these systems yield the major contribution to conduction properties.~\cite{SaitoDD98} Then the Hamiltonian describing our system can be written in the tight-binding approach as follows:
\begin{equation}
\Ham_0 = \sum_{i} \epsilon_{2p} c_{i}^\dagger c_{i}^{\phantom{\dagger}} +
\sum_{\lab i,j \rab} t_{ij}^0 c_{i}^\dagger c_{j}^{\phantom{\dagger}},
\label{eq:Ham}
\end{equation}
where the indices $\lab i,j \rab$ indicate nearest neighbor atom sites and the summation is extended over all the points in the lattice.
The onsite energies are $\epsilon_{2p}$ and $t_{ij}^0$ are the hopping parameters between neighboring sites.
We will shift our energy scale in order to have vanishing on-site energies, setting $\epsilon_{2p} = 0$.
The parameter $t_{ij}^0$ is given by the hopping between nearest neighbor $p_z$ orbitals, $V_{pp}^\pi$.

The above Hamiltonian in real space needs to be modified in order to incorporate the effects we are interested in: (A) a parallel magnetic field, (B) the curvature of the carbon nanotubes, (C) the spin-orbit interaction. In the following we will therefore find the modified hopping parameters providing these effects.

From now on we shall be using two sets of coordinates to describe our system.
The first, $(x_i,y_i,z_i)$, denotes the local direction of the orbitals at the atom $i$ of the nanotube. The direction $x_i$ is tangent to the circumference of the tube, $y_i$ parallel to the nanotube axis, and $z_i$ is always perpendicular to the nanotube surface  as seen in Fig.~\ref{fig:coord}a.
The second system, $(X,Y,Z)$ accounts for the global cartesian coordinates of the nanotube, with the $Y$-axis coinciding with the tube axis. In the global cylindrical coordinates $(\radio,\azim,Y)$ the angle $\azim$ is measured from the positive $Z$-axis and $\radio$ is the nanotube radius.
Therefore when we consider the atomic orbitals, the $\pi$ orbital is in the $z_i$ direction, perpendicular to the nanotube surface and the magnetic field will be applied along the $Y$-direction.

\subsection{Parallel magnetic field}\label{ssec_model:MagnField}

In the presence of a magnetic field the Hamiltonian describing our system is obtained by the substitution $-\ii\hbar\nabla \rightarrow -\ii\hbar\nabla - \frac{e}{c}\vect{A}$, where $\vect{A}$ is the vector potential of the magnetic field $\vect{B}$. As shown by Luttinger,~\cite{Luttinger51} the effect of the magnetic field in the Schr\"odinger equation is greatly simplified if the Bloch functions are modified by a phase factor $\exp{\lrb \ii \frac{e}{\hbar}\int_{\vect{R}_i}^{\vect{r}}\vect{A}(\vect{r'})\textrm{d}\vect{r'} \rrb}$.~\cite{limitsPeierls} Therefore each matrix element in the Hamiltonian is obtained by multiplication of the corresponding matrix element in zero magnetic field by a phase factor:
\begin{equation}
 t_{ij}^0 \lrb \vect{B} \rrb = t_{ij}^0 \lrb 0\rrb  e^{\frac{\ii e}{\hbar}\int^{\vect{R}_i}_{\vect{R}_j}\vect{A}(\vect{r})\textrm{d}\vect{r}},
\end{equation}
known as Peierls phase factor,~\cite{Peierls33} where the integral is taken along a line joining the sites $i$ and $j$.

In the case of a uniform static magnetic field in the direction of the nanotube axis, $\vect{B}=(0,B_0,0)$, the vector potential can be given by
$\vect{A} = \frac{B_0}{2}\lrb -Z,0,X \rrb$,
or $\lrb 0,\,\frac{1}{2} B_0 r,\, 0\rrb$ in cylindrical coordinates, where we have taken the Coulomb or transverse gauge. With this choice of gauge the Peierls phase factor is given by
\begin{equation}
 e^{\frac{\ii e}{2\hbar}\int^{\azim_i}_{\azim_j}B_0 r^2 \textrm{d}\azim} =
 e^{\ii\frac{\phi}{\phi_0}\lrb\azim_i-\azim_j \rrb},
\end{equation}
where $\phi$ is the magnetic flux in the cross section of the nanotube perpendicular to its axis, $\phi_0=h/e$ is the flux quantum, and $\theta_i-\theta_j$ is the angular difference between the azimuthal coordinates of sites $i$ and $j$.

The Zeeman effect appearing in the presence of a magnetic field is taken into account by including in our Hamiltonian the term $ \Ham_Z=\frac{g_e}{2} \mu_B \vect{B}\cdot\vec{\sigma}$, where $g_e$ is the electronic g-factor, $\mu_B=\frac{\ee\hbar}{2m}$ is the Bohr magneton and $\vec{\sigma}$ is Pauli's spin matrix vector.
With the magnetic field directed along the $Y$-axis, only the Pauli matrix $\sigma_Y$ remains, which conserves the spin in the $Y$-direction with eigenvalues $\pm 1$ for parallel/antiparallel spins in this direction.
Expressing the magnetic field $B_0$ in terms of the magnetic flux and with the g-factor $g_e$ set to $2$, the Zeeman term is then
\begin{equation}
 \Ham_Z = \frac{\phi}{\phi_0}\frac{\hbar^2}{m \radio^2}\sigma_Y,
\end{equation}
giving a corresponding shift of $\pm \frac{\phi}{\phi_0}\frac{\hbar^2}{m \radio^2}$ for spins up/down in the $Y$-direction.

\subsection{Curvature}\label{ssec_model:curv}

\begin{figure}[h!]
\centerline{\includegraphics[width=.5\linewidth]{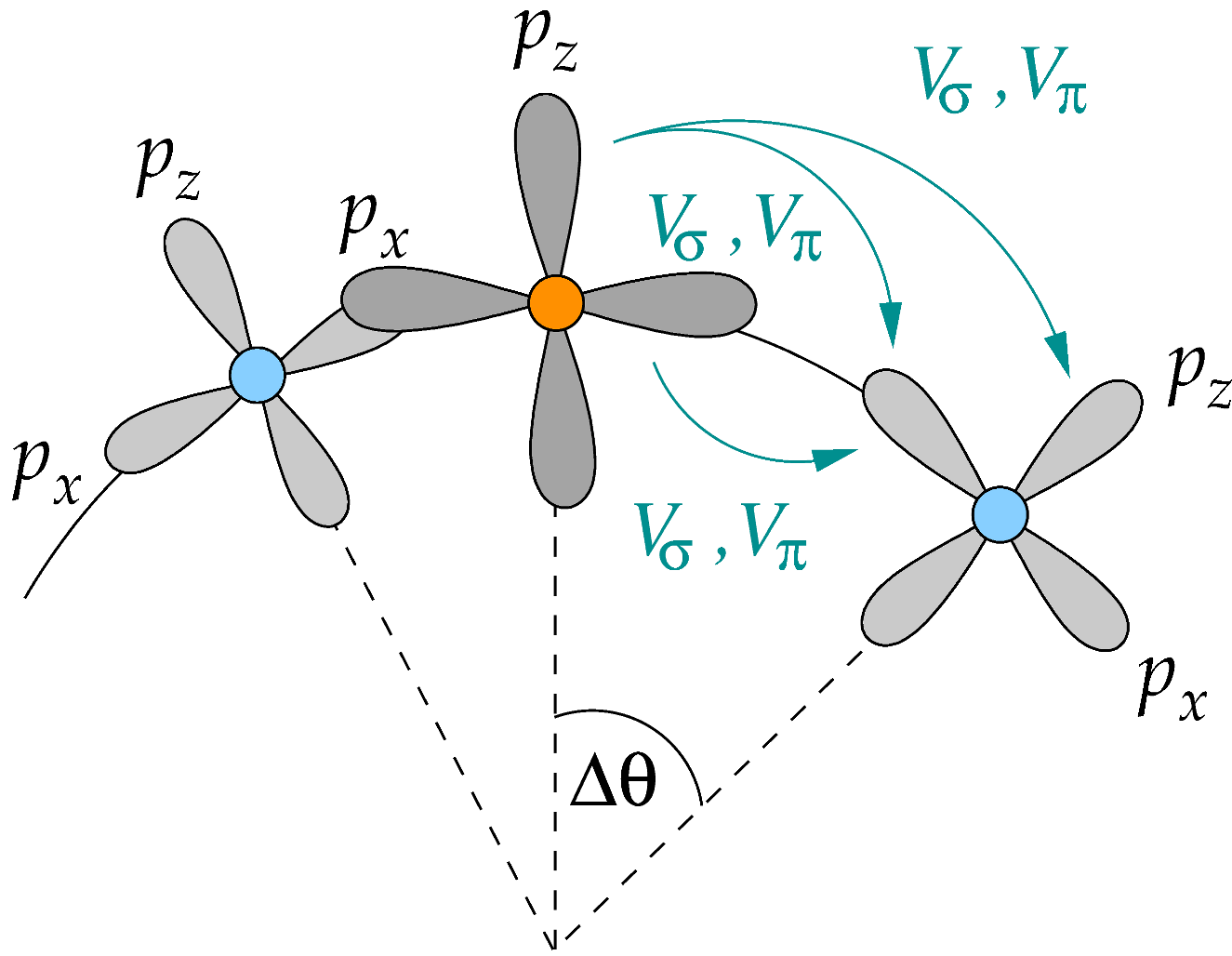}}
\caption{\label{fig:curv}
Schematic representation of the effect of the curvature of a nanotube on the transfer integrals between the orbitals of nearest neighbor atoms. The $p_z$ orbitals are not parallel anymore, so that a mixing with the orbitals building the $\sigma$ bonds takes place.
}
\end{figure}

Due to the curvature of the nanotube lattice, the $2p_z$ orbitals which are parallel to each other in the graphene are not parallel anymore (see Fig.~\ref{fig:curv}). This new interplay of the chemical bonds has to be taken into account, and we have to recalculate the hopping between $p_z$ orbitals accordingly. This hopping will now have not only components due to orbitals normal to the atomic plane, but also due to tangent ones. The hopping between two neighboring $p_z$ orbitals at sites $i$ and $j$ is then
\begin{equation}
\rbra{z_i}\Ham\rket{z_j}  =  V_{pp}^\pi\; \vect{n}^\perp(z_i)\cdot\vect{n}^\perp(z_j)
+ V_{pp}^\sigma\; \vect{n}^\parallel(z_i)\cdot\vect{n}^\parallel(z_j)
\end{equation}
where $\vect{n}^\perp(z_i)$ and $\vect{n}^\parallel(z_i)$ are the normal and tangential components of a unit vector in the direction of the $p_z$ orbital at site $i$, $\rket{z_i}$, and
$V_{pp}^\sigma$ and $V_{pp}^\pi$ are the transfer integrals giving rise to $\sigma$ and $\pi$ orbitals in flat 2D graphene, respectively.
We use throughout the paper round brackets to indicate states in the absence of spin-orbit interaction and we will leave the normal bracket notation for the states that include this perturbation, as introduced in Ref.~\onlinecite{Ando00}.
The parallel component of $\vect{n}(\alpha_i)$ is its projection along $\vect{R}_{ji}$ given by
$\vect{n}^\parallel(\alpha_i)  =  \frac{\vect{n}(\alpha_i)\cdot\vect{R}_{ji}}{\abso{R_{ji}}^2}\,\vect{R}_{ji}$, and $\vect{n}^\perp(\alpha_i) =  \vect{n}(\alpha_i) -  \vect{n}^\parallel(\alpha_i)$, where  $\vect{R}_{ji}=\vect{R}_j-\vect{R}_i$ is the vector connecting the two sites.
Therefore
\begin{equation}
\begin{split}
\rbra{z_i}\,\Ham\,\rket{z_j}  &=  V_{pp}^\pi\; \vect{n}(z_i)\cdot\vect{n}(z_j)\\
&+ (V_{pp}^\sigma - V_{pp}^\pi) \frac{(\vect{n}(z_i)\cdot\vect{R}_{ji}) (\vect{n}(z_j)\cdot\vect{R}_{ji})}{\abso{R_{ji}}^2}.
\end{split}
\label{eq:oti}
\end{equation}
For the case of $p_z$ orbitals this means that
\begin{equation}
\begin{split}
\rbra{z_i}\Ham\rket{z_j} &= V_{pp}^\pi \cos{\lrb\theta_i -\theta_j\rrb}\\
    &- (V_{pp}^\sigma - V_{pp}^\pi) \frac{\radio^2}{\aC^2} \lsb 1 -\cos{\lrb\theta_i-\theta_j\rrb}\rsb^2,
\end{split}
\end{equation}
where $\aC=\abso{R_{ji}}$ is the distance between neighboring atoms.

\subsection{Spin-orbit coupling (\SOIns)}\label{ssec_model:SOI}

Carbon $2p$ orbitals present a weak spin-orbit interaction (\SOI).~\cite{MinHSSKM06} Because the effect of this interaction is small we shall treat it as a perturbation, and consider its effects to the lowest order.

We follow here the method used by Ando,~\cite{Ando00} and adapt it to the case of spins in the direction of the parallel magnetic field. First we have to calculate the modification of the $2p_z$ atomic orbitals once the \SOI is included. The Hamiltonian describing this interaction is given at the atomic level by~\cite{Winkler03}
\begin{equation}
 \Ham_{so} = \frac{\hbar}{4 m^2 c^2} \lrb \nabla V \times \vect{p} \rrb\cdot \vec{\sigma},
\label{eq:Ando_Hso}
\end{equation}
where $V (\vect{r})$ is the atomic potential, $m$ is the free-electron mass and $\vect{p}$ is the momentum operator

As described in Appendix~\ref{sec_appA} we include the \SOI Hamiltonian as a perturbation and obtain to first order the modified $p_z$ orbitals, $\ket{z_j s}$ with $s=\uparrow_y,\downarrow_y$, denoting the spin up/down in the $Y$-direction. The perturbed orbital $\ket{z_j s}$ does not have anymore a well defined spin and the spin in the axial direction $s$ indicates merely the main character of the state. Moreover $\ket{z_j s}$ contains also an admixture of the $p$ orbitals building the $\sigma$ bonds.

The transfer integrals for these perturbed states are now given by
\begin{widetext}
\begin{equation}
\begin{split}
 t_{\uparrow\uparrow} &= \bra{z_i\uparrow_y}\, \Ham\, \ket{z_j\uparrow_y}
 \simeq  \; V_{pp}^\pi \cos{\lrb\theta_i-\theta_j\rrb} - (V_{pp}^\sigma - V_{pp}^\pi) \frac{\radio^2}{\aC^2} \lsb\cos{\lrb\theta_i-\theta_j\rrb} - 1\rsb^2\\
&+ 2\ii\delta\lcb V_{pp}^\pi \sin{\lrb\theta_i-\theta_j\rrb} + (V_{pp}^\sigma - V_{pp}^\pi) \frac{\radio^2}{\aC^2} \sin{\lrb\theta_i-\theta_j\rrb} \lsb 1 - \cos{\lrb\theta_i-\theta_j\rrb}\rsb\rcb,\\
 t_{\downarrow\downarrow} &= \bra{z_i\downarrow_y}\, \Ham\, \ket{z_j\downarrow_y}
 =  \bra{z_i\uparrow_y}\, \Ham\, \ket{z_j\uparrow_y}^*,\\
 t_{\uparrow\downarrow} &= \bra{z_i\uparrow_y}\, \Ham\, \ket{z_j\downarrow_y}
 \simeq   -\delta\;\lrb e^{-\ii\theta_j} + e^{-\ii\theta_i} \rrb (V_{pp}^\sigma - V_{pp}^\pi) \frac{\radio Y_{ji}}{\aC^2} \lsb \cos{\lrb\theta_i-\theta_j\rrb - 1}\rsb,\\
 t_{\downarrow\uparrow} &= \bra{z_i\downarrow_y}\, \Ham\, \ket{z_j\uparrow_y}
 =  -\bra{z_i\uparrow_y}\, \Ham\, \ket{z_j\downarrow_y}^*,
\end{split}
\label{eq:newt}
\end{equation}
\end{widetext}
where $Y_{ji}$ is the distance between the interacting atoms projected in the axial direction and $\delta$ is a dimensionless parameter measuring the \SOI strength as defined in Appendix~\ref{sec_appA}.

With these recalculated hopping parameters we can build our Hamiltonian in real space and numerically compute the transport properties of our systems, \cf Sec.~\ref{sec_numerical_transp}. In order to gain a deeper insight in the results, it is important though to approach the problem analytically as well.
Hence in the coming section we derive from Eq.~(\ref{eq:Ham}) and with Eq.~(\ref{eq:newt}) a low energy Hamiltonian in $\vect{k}$-space.
In this way we will not only be able to understand the observed numerical features but also to make analytical predictions and save computational costs.

\subsection{Low energy Hamiltonian}\label{ssec_model:dispersion}

The Hamiltonian of carbon nanotubes can be written in the basis of the Bloch functions for the $A$ and $B$ sublattices in the same way as for graphene:~\cite{Wallace47} $\Ham(\vect{k})=\lrb\begin{smallmatrix}
                       0  & \Ham_{AB}^{\phantom{\dagger}}(\vect{k})\\
                       \Ham_{AB}^\dagger(\vect{k}) &              0
                      \end{smallmatrix}\rrb$ with $\Ham_{AB}^{\phantom{\dagger}}(\vect{k}) = \sum_{l=1}^3 t_l \ee^{\ii \vect{k}\cdot\vect{d}_l}$. As shown in
Fig.~\ref{fig:coord}c, the vectors connecting neighboring atoms are denoted by $\vect{d}_l$.
As the spin degeneracy is broken, the transfer integral $t_l $ in $\Ham_{AB}^{\phantom{\dagger}}(\vect{k})$ becomes now a 2x2 matrix:
$t_l= \lrb \begin{smallmatrix} t_{\uparrow\uparrow}(\vect{d}_l) & t_{\uparrow\downarrow}(\vect{d}_l)\\ t_{\downarrow\uparrow}(\vect{d}_l) & t_{\downarrow\downarrow}(\vect{d}_l) \end{smallmatrix}\rrb$. These matrix elements are evaluated in Appendix~\ref{sec_appB} explicitly in the limit of large radii, so that $\aC/\radio$ is a small quantity, and in the limit of low energies, such that an expansion of the Hamiltonian around the $\Kone$ and $\Ktwo$ points of the reciprocal lattice (see Fig.~\ref{fig:coord}d) is justified. The matrix block $\Ham_{AB}^{\phantom{\dagger}}(\vect{k})$ becomes in this latter limit $\Ham_{AB}^\tau(\vectGr{\kappa})$, valid for small vectors $\vectGr{\kappa}$ around the Dirac point $\tau\Kone$ with $\tau=+1$ for $\Kone$ and $\tau=-1$ for $\Ktwo$ ($\vect{k}=\tau\Kone+\vectGr{\kappa}$).

With the definitions $\Delta k_\perp^c = \frac{\aC}{4 R^2}  \tau p \cos{3\eta} $, $\Delta k_\parallel^c = -\frac{\aC}{4 R^2} p' \sin{3\eta}$, $\Delta k_\perp^{SO}=\frac{2\delta p}{R}$, and $\Delta_{flip}=\frac{\delta}{4 R} \frac{V_{pp}^\sigma - V_{pp}^\pi}{V_{pp}^\pi}$, with $p= 1+\frac{3}{8}\frac{V_{pp}^\sigma - V_{pp}^\pi}{V_{pp}^\pi}$, $p'= 1+\frac{5}{8}\frac{V_{pp}^\sigma - V_{pp}^\pi}{V_{pp}^\pi}$,  we can write $\Ham_{AB}^\tau(\vectGr{\kappa})$ as
\begin{equation}
\begin{split}
&\Ham_{AB}^\tau(\vectGr{\kappa}) =
  \hbar \upsilon_F \ee^{-\ii \tau \eta} \Bigg\{
\!\!\!\lrb \tau \kappa_\perp + \tau \Delta k_\perp^c\rrb
- \ii\!\lrb \kappa_\parallel + \tau \Delta k_\parallel^c\rrb \!\!\!\Bigg\}
  \mathbbm{1}\\
&- \hbar \upsilon_F \ee^{-\ii \tau \eta}\Bigg\{ \tau \Delta k_\perp^{SO}\Bigg\}
  \sigma_Y
- \hbar \upsilon_F \ee^{-\ii \tau \eta}\Bigg\{ \Delta_{flip} \Bigg\}
  \ii \sigma_{\tilde{x}_{ij}},
\end{split}
\end{equation}
where $\upsilon_F=\frac{3\pi\aC V_{pp}^\pi}{h}\approx 8.6\cdot10^{5}\textrm{m}/\textrm{s}$ is the group velocity at the Fermi points.
Notice that the spin-flipping term is actually a rotation into the local coordinate system $\lrb\begin{smallmatrix} 0&\ee^{-\ii\tilde{\theta}}\\ -\ee^{\ii\tilde{\theta}}&0 \end{smallmatrix}\rrb
= \ii\lrb \cos{\tilde{\theta}}\sigma_X - \sin{\tilde{\theta}}\sigma_Z\rrb = \ii \sigma_{\tilde{x}_{ij}}$, $\tilde{x}_{ij}$ being the direction tangential to the tube at the middle point of the bond between $i$ and $j$ atoms.

The eigenvalues of $\Ham^\tau$ are then
\begin{equation}
\begin{split}
  E_\pm &\lrb\tau,\sigma,\kappa_\perp,\kappa_\parallel\rrb =  \pm \hbar \upsilon_F \lcb (\tau \kappa_\perp + \tau \Delta k_\perp^c)^2 + (\Delta k_\perp^{SO})^2\rnothing\\
&+ (\kappa_\parallel +\tau \Delta k_\parallel^c)^2 + \Delta_{flip}^2 \\
&+ 2 \sigma \lsb\lrb (\Delta k_\perp^{SO})^2 + (\kappa_\parallel + \tau \Delta k_\parallel^c)^2\rrb  \Delta_{flip}^2\rnothing\\
&+ \lnothing\lnothing(\tau \kappa_\perp + \tau \Delta k_\perp^c)^2 (\Delta k_\perp^{SO})^2\rsb^{1/2}\rcb^{1/2},\\
\end{split}
\end{equation}
with $\sigma=\pm 1$ referring to the spin, so that we could merge all four eigenvalues into this expression.

If we neglect the term allowing for spin flip ($\Delta_{flip}$), which is small in comparison to other contributions, we have
\begin{equation}
 E_\pm \lrb\tau,\sigma,\kappa'_\perp,\kappa'_\parallel\rrb = \pm \hbar \upsilon_F \sqrt{\lrb \kappa'_\perp\rrb^2 + \lrb \kappa'_\parallel \rrb^2},
\label{eq:dr_general}
\end{equation}
equivalent to the dispersion relation for the planar graphene, but with the definitions
\begin{equation}
\begin{split}
 \kappa'_\perp &= \kappa_\perp + \Delta k_\perp^c + \sigma \Delta k_\perp^{SO},\\
 \kappa'_\parallel &= \kappa_\parallel + \tau \Delta k_\parallel^c.
\end{split}
\label{eq:kappaprime}
\end{equation}
Neglecting the spin-flipping terms we recover a well-defined spin in our problem, so that we can distinguish between spin-up and spin-down eigenfunctions.

The magnetic field modifies the dispersion relation further via the Aharonov-Bohm term, which changes  $\kappa_\perp$ by $\frac{1}{R}\frac{\phi}{\phi_0}$, so that $\kappa'_\perp$ in Eq.~(\ref{eq:kappaprime}) becomes
\begin{equation}
 \kappa'_\perp = \kappa_\perp + \Delta k_\perp^c + \sigma \Delta k_\perp^{SO} + \frac{1}{R}\frac{\phi}{\phi_0}.
\label{eq:kappaprimeTotal}
\end{equation}

The contribution of the spin-orbit interaction on the diagonal of the matrix blocks can be also interpreted as an effective magnetic field and we can write
$\Delta k_\perp^{SO}=\frac{1}{R}\frac{\phi_{SO}}{\phi_0}$ with $\phi_{SO}/\phi_0=2\delta p$.
The other contributions, $\Delta k_\perp^c$ and $\Delta k_\parallel^c$, arise from the consideration of the curvature in the structure of the nanotubes, as it can be seen in Ref.~\onlinecite{MarganskadVJSG11}.
The Zeeman effect modifies further the dispersion relation given in Eq.~(\ref{eq:dr_general}) by introducing a spin-dependent shift:
\begin{equation}
 E_\pm \lrb\tau,\sigma,\kappa'_\perp,\kappa'_\parallel\rrb = \pm \hbar \upsilon_F \sqrt{\lrb \kappa'_\perp\rrb^2 + \lrb \kappa'_\parallel \rrb^2}
 + \sigma \mu_B \abso{\vect{B}},
\label{eq:ana_final}
\end{equation}
that is, levels of up (down) spins are correspondingly shifted up (down) in energy.
Ref.~\onlinecite{IzumidaSS09} introduces an additional effective valley-dependent Zeeman term which leads to a small energy splitting in the Dirac cones and introduces a further electron-hole asymmetry. The magnitude of this shift only slightly modifies this spectrum, and will therefore not be considered in the following calculations.

\section{Finite size nanotubes}\label{sec_model:dispersion_BC}

Eq.~(\ref{eq:ana_final}) allows us to calculate analytically the spectrum of \CNTs in a parallel magnetic field, including \SOI and the Zeeman effect,
when in addition proper boundary conditions are imposed that yield the appropriate quantization of the vector $\vectGr{\kappa}$.

First, as a \CNT is obtained by rolling a graphene layer into a tube, in the angular direction its wave functions always obey periodic boundary conditions:
$\psi\lrb\chiralvec\rrb = \exp\lrb\ii\vect{k}\cdot\chiralvec\rrb\psi\lrb 0\rrb = \ee^{\ii2\pi n}\psi\lrb 0\rrb$.
This boundary condition leads to a quantization of the transverse component of $\vect{k}$, $k_\perp  = \frac{2\pi}{\abso{\chiralvec}}l_\perp$,
with $l_\perp$ being a positive integer. We can obtain the energy spectrum of infinitely long carbon nanotubes from the dispersion relation of graphene by restricting the allowed values of $\vect{k}$ by the quantization condition given above. The quantized $\kappa_\perp$ are then obtained from the relation $\kappa_\perp = k_\perp-\tau\Kone_\perp$, which together with Eq.(\ref{eq:kappaprimeTotal}) determines the quantized $\kappa'_\perp$ values.

To obtain the quantization of $\kappa_\parallel$, and therefore of $\kappa'_\parallel$,  for \CNTs of finite length we have to impose open boundary conditions in the axial direction, that is, we impose that our wave function vanishes at the ends of the tube. To this extent we have to consider the underlying structure of the nanotube. Here we distinguish between two cases: i) armchair nanotubes, where each end contains the same number of atoms from the $A$ and $B$ sublattices and ii) zigzag and chiral nanotubes, where each end is formed predominantly by one type of atom. Because the same boundary conditions apply for zigzag nanotubes and any other chiral \CNTns,~\cite{AkhmerovB08} provided the nanotube edge is a so-called minimal boundary (there are no atoms with only one neighbor), we shall derive them for the zigzag case. In zigzag \CNTs one end is entirely composed of $A$ sublattice atoms and the opposite end is formed only by $B$ sublattice atoms.

\textit{Zigzag nanotubes:} The open boundary conditions in this case imply that the wave function must vanish on one end at the missing A atoms and on the other end at the missing B atoms. We therefore impose
\begin{equation}
 \begin{split}
  &\psi_{\tau A}\lrb (x_\perp,x_\parallel=L), E_\pm\rrb =0,\\
  &\psi_{\tau B}\lrb (x_\perp,x_\parallel=0), E_\pm\rrb =0,\\
 \end{split}
 \label{eq:BCzz}
\end{equation}
where the wave functions $\psi_{\tau A}$, $\psi_{\tau B}$ are the sublattice components of $\psi_{\tau}\lrb \vect{r}, E_\pm\rrb$, the eigenstate at a given energy $E(\kappa'_\perp,\kappa'_\parallel)$ and at the Fermi point $\tau\Kone$. Because the solutions with $\kappa'_\parallel$ and $-\kappa'_\parallel$ are degenerate in energy, as can be seen from the energy dispersion, Eq.~(\ref{eq:ana_final}), the eigenstate satisfying Eq.~(\ref{eq:BCzz}) will be a linear combination
of the eigenstate for $\kappa'_\parallel$ and the one for $-\kappa'_\parallel$, or equivalently the one for $\kappa_\parallel$ and $-\kappa_\parallel-2\tau\Delta k_\parallel^c$.
As shown in Ref.~\onlinecite{MarganskadVJSG11} the condition to be fulfilled by the $\vectGr{\kappa}$ values is then
\begin{equation}
 \frac{\tau \kappa'_\perp + \ii\kappa'_\parallel}{\tau \kappa'_\perp - \ii\kappa'_\parallel} = \ee^{\ii 2\kappa'_\parallel L}.
\label{eq:BCzz_sol}
\end{equation}
This equation has both extended and localized solutions and a trivial solution for $\kappa'_\parallel=0$.
If $\kappa'_\parallel$ is real the condition given by Eq.~(\ref{eq:BCzz_sol}) becomes
\begin{equation}
 \tau \kappa'_\perp = \kappa'_\parallel \cot{\kappa'_\parallel L},
\label{eq:BCzz_real}
\end{equation}
which represents the extended solutions. Otherwise if $\kappa'_\parallel$ is a pure imaginary number Eq.~(\ref{eq:BCzz_sol}) becomes
\begin{equation}
 \tau \kappa'_\perp = \ImP{(\kappa'_\parallel)} \coth{\lrb\ImP{(\kappa'_\parallel)} L\rrb}.
\label{eq:BCzz_im}
\end{equation}
As $\kappa'_\parallel$ is imaginary, the solution to this equation describes evanescent waves.
Eq.~(\ref{eq:BCzz_im}) has two non-trivial solutions:
$\kappa'_\perp\geq 1/L$ for the $\Kone$-point and $\kappa'_\perp \leq -1/L$ for the $\Ktwo$-cone.

For $\kappa'_\perp = 0$, Eq.~(\ref{eq:BCzz_sol}) yields as allowed values of $\kappa'_\parallel$:
\begin{equation}
 \kappa'_\parallel= \lrb 2n +1\rrb \frac{\pi}{2L},
\label{eq:BCzzkpar}
\end{equation}
with $n\in\mathbb{Z}$.

\begin{figure}[h!]
\centerline{\includegraphics[width=\linewidth]{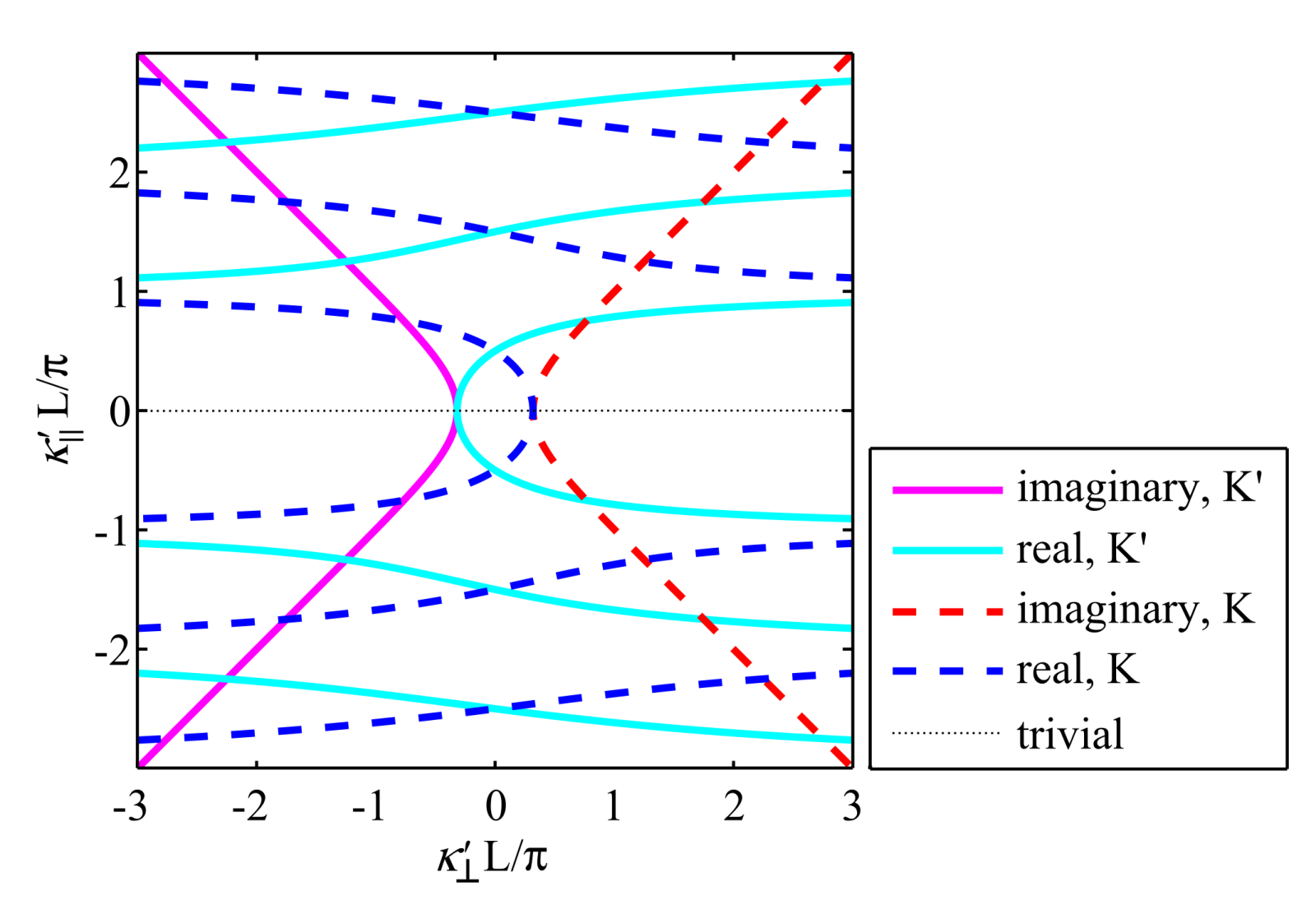}}
\caption{\label{fig:allowedk}
Allowed $\kappa'_\perp$ and $\kappa'_\parallel$ values for a zigzag-like \CNT fulfilling the boundary condition given by Eq.~(\ref{eq:BCzz_sol}). At $\kappa'_\perp=0$, as expected, $\kappa'_\parallel \frac{L}{\pi} = n+\frac{1}{2}$ with $n\in\mathbb{Z}$ and when including the spin the Kramers degeneracy holds.
}
\end{figure}

The allowed solutions of Eq.~(\ref{eq:BCzz_real}) and Eq.~(\ref{eq:BCzz_im}) are plotted in Fig.~\ref{fig:allowedk}.

\begin{figure}[h!]
\centerline{\includegraphics[width=\linewidth]{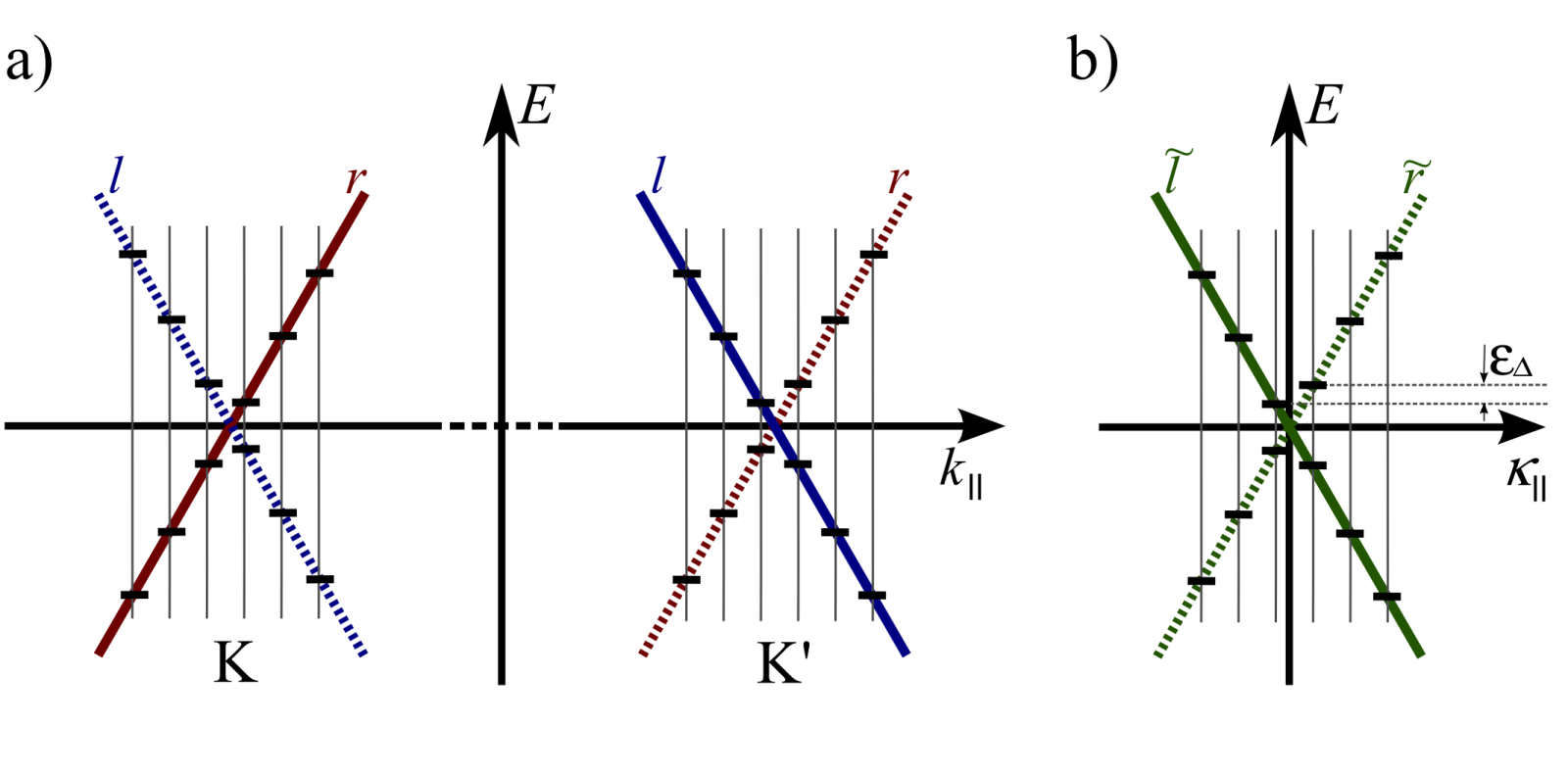}}
  \vspace{-10pt}
\caption{
Linear dispersion relation of finite size armchair carbon nanotubes with open boundary conditions. a) The band structure of an armchair \CNT with periodic boundary conditions around the two inequivalent Fermi points, $\Kone$ and $\Ktwo$.
For finite-length tubes only certain values of $\kappa'_\parallel$ are allowed which are here indicated by the sketched vertical cuts.
Branches of left movers are indicated by $l$  whereas $r$ indicate those of right movers.
b) These four branches can be combined into the band structure shown on the right, which represents the solutions for
the open boundary problem. We have chosen explicitly the case where $\kappa_\perp=0$ and the allowed $\kappa_\parallel$ do
not lie symmetrically around the Fermi points, in order to easily observe the bands of equal energy. Including \SOI and the
magnetic field in $\kappa'_\perp$ leads to shifted cuts of the Dirac cones and we would have the same picture as here but instead with parabolas.
}\label{fig:BZcones_ACcombination}
\end{figure}

\textit{Armchair nanotubes:} The infinite armchair
\CNT will be always metallic in the $\pi$-orbital approximation, as its chiral angle of $30^\circ$ assures the cut of the $\Kone$ and $\Ktwo$ points.
Bloch waves on branches of the linear dispersion relation with positive (negative) group velocities, $\upsilon_F$, are called right (left) movers, as seen in Fig.~\ref{fig:BZcones_ACcombination}.

The armchair case was not discussed in Ref.~\onlinecite{MarganskadVJSG11}. Here we extend the treatment of Ref.~\onlinecite{MayrhoferG08} to include \SOI and magnetic field effects.
To impose the open boundary conditions we have to build once more linear combinations of Bloch waves (in the expansion around the Fermi points) which are
degenerate in energy.
In the presence of an off-set $\Delta$, which accounts for the case that $\abso{\R{K}_\parallel}=\frac{\pi}{L}\lrb m-\Delta\rrb$ with $m$ being a positive integer, the branches that are equivalent in energy are the $r$($l$)-band at $\Kone$ and the $l$($r$)-band at $\Ktwo$.
The linear combinations of wavefunctions are then:
\begin{equation}
 \begin{split}
  &\psi_{\tilde{r}}\lrb \vect{r}, \vectGr{\kappa}\rrb = \frac{1}{\sqrt{2}}\lrb \Phi_{r}^\Ktwo(\vect{r}, (\kappa_\perp,\kappa_\parallel)) - \Phi_{l}^\Kone(\vect{r}, (\kappa_\perp,-\kappa_\parallel))\rrb,\\
  &\psi_{\tilde{l}}\lrb \vect{r}, \vectGr{\kappa}\rrb = \frac{1}{\sqrt{2}}\lrb \Phi_{l}^\Ktwo(\vect{r}, (\kappa_\perp,\kappa_\parallel)) - \Phi_{r}^\Kone(\vect{r}, (\kappa_\perp,-\kappa_\parallel))\rrb,\\
 \end{split}
\end{equation}
where $\Phi_{l/r}^{\tau\Kone}$ are the Bloch wave functions for the $l/r$ branches. We have therefore to mix states which belong to different Dirac cones.

An analysis of the wave functions $\psi_{\pm}$ reveals that they are proportional to $\sin{\lrb \lrb \R{K}_\parallel +\kappa_\parallel\rrb x_\parallel\rrb}$. The open boundary conditions on these wave functions are given by~\cite{length}
\begin{equation}
\begin{split}
 \psi_{\tilde{r}/\tilde{l}}\lrb(x_\perp,x_\parallel=L);\vectGr{\kappa}\rrb&=0, \\
 \psi_{\tilde{r}/\tilde{l}}\lrb(x_\perp,x_\parallel=0);\vectGr{\kappa}\rrb&=0,
\end{split}
\label{eq:BCac}
\end{equation}
and require the wave functions to vanish on atom sites belonging to both sublattices (as opposed to the case of zigzag nanotubes).
They are then fulfilled if $\lrb\R{K}_\parallel+\kappa_\parallel\rrb L = \pi n$, with $n\in\mathbb{Z}$. That is,
\begin{equation}
 \kappa_\parallel = \frac{\pi}{L} \lrb n_\kappa +\Delta \rrb, \text{ with } n_\kappa \in\mathbb{Z}.
\end{equation}
A value of $\Delta<1$ different from $0$ or $\pm\frac{1}{2}$ causes a mismatch $\epsilon_\Delta=\hbar\upsilon_F\frac{\pi}{L}\lrb 2\Delta-1\rrb$ between the $\tilde{r}$ and $\tilde{l}$ branches as
seen in Fig.~\ref{fig:BZcones_ACcombination}. One of the bands will therefore be energetically favored if an electron is added to the tube.

To include the effects of \SOI and an applied magnetic field it suffices to note that the Bloch wave functions building up $\psi_{\tilde{r}/\tilde{l}}$ for a fixed spin $\sigma$ are still degenerate when considering Eq.~(\ref{eq:ana_final}). Indeed $\kappa'_\parallel$ includes a valley-dependent shift, so that the Bloch wave functions $\Phi_{r/l}^\Ktwo(\vect{r}, (\kappa_\perp,\kappa_\parallel),\sigma)$ and $\Phi_{l}^\Kone(\vect{r}, (\kappa_\perp,-\kappa_\parallel),\sigma)$ are characterized by $\kappa'_\parallel=\kappa_\parallel-\Delta k_\parallel^c$ and $\kappa'_\parallel=-\kappa_\parallel+\Delta k_\parallel^c$ and are therefore degenerate in energy.

The allowed values for $\kappa'_\parallel$ in the case of armchair \CNTs are all real and no imaginary solutions are possible. Thus all states building the energy spectrum of armchair tubes will correspond to extended states.

\section{Transport properties in a parallel magnetic field}\label{sec_numerical_transp}

\begin{figure}[h!]
\centerline{\includegraphics[width=\linewidth]{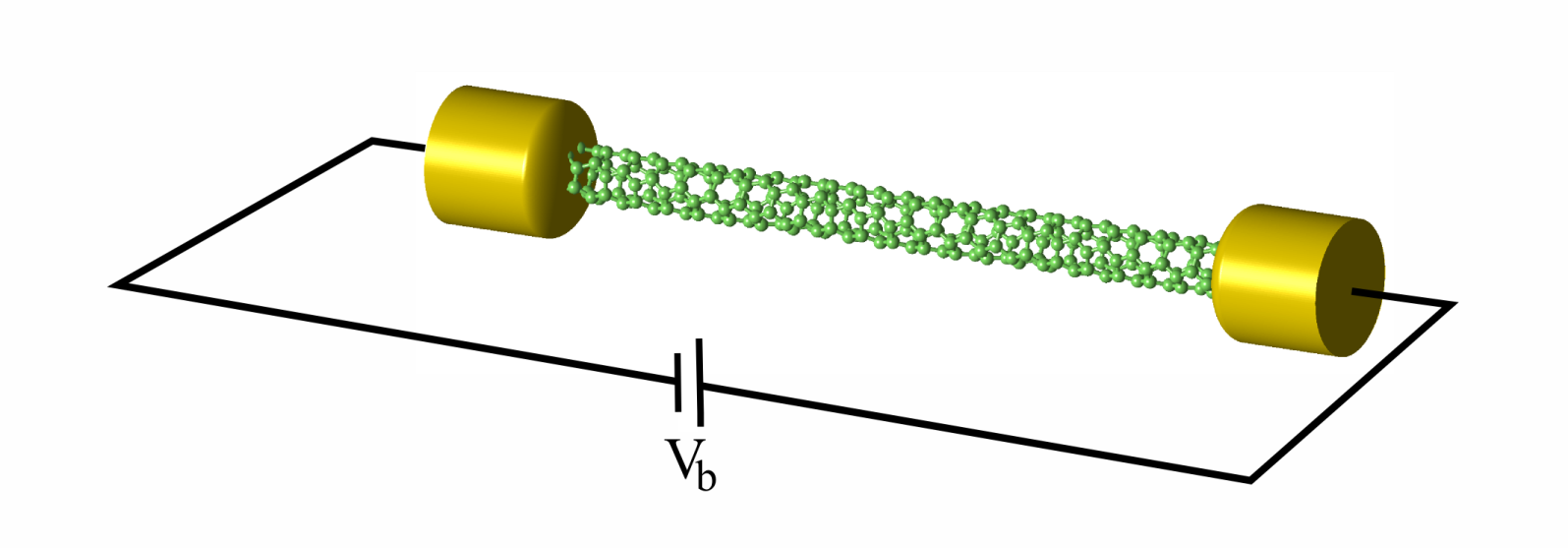}}
\caption{\label{fig:scheme}
Schematical drawing of the set-up of our system, where a suspended \CNT connects the source and drain leads, between which a bias voltage $V_b$ is applied.
}
\end{figure}

The transport through the systems we studied in the previous sections is calculated by connecting the \CNTs to metallic leads, as schematically shown in Fig.~\ref{fig:scheme}. Nanotubes of various chiralities and lengths are considered. Their transport characteristics are obtained from the transmission values gained by applying the Landauer formula, using Green function
techniques.~\cite{CunibertiGG02} In particular, we determine the elastic linear response conductance via the Fisher-Lee formula for
the quantum mechanical transmission~\cite{FisherL81}:
$G=\frac{2e^2}{h} \textrm{Tr}\lcb \vect{\Gamma}_{\rm L} \,
\vect{{\cal G}}  \, \vect{\Gamma}_{\rm R} \, \vect{{\cal G}}^\dagger
\rcb,$ where $\vect{\Gamma_{\rm L/R}} = \ii ( \vect{\Sigma}_{\rm
L/R} - \vect{\Sigma}_{\rm L/R}^\dagger )$, $\vect{\Sigma}_{\rm L/R}$
is the self energy associated to the left or right lead respectively, and $\vect{{\cal G}}$ is the Green function of the central region
dressed by the electrodes, $\vect{{\cal G}}=\lrb E - \Ham - \vect{\Sigma}_{\rm L} - \vect{\Sigma}_{\rm R}\rrb^{-1}$.
For simulating bulk metal electrodes we will consider wide band leads, \ie $\vect{\Sigma}_{{WB}} (E) \cong
-\ii \,\mathrm{Im}\vect{\Sigma}(E=E_{\textrm F}$).

We will now show in detail the spectrum and transport characteristics of a finite zigzag nanotube of chirality (9,0) with a length of hundred unit cells (about 42 nm). In order to see the influence in transport of the different effects considered, we will include them gradually. Moreover we will consider a rather large value of $\delta$ to observe clearly the effects of the spin-orbit interaction. All other parameters used for the following calculations are taken from Ref.~\onlinecite{TomanekL88}.

\begin{figure}[h!]
\centerline{\includegraphics[width=\linewidth]{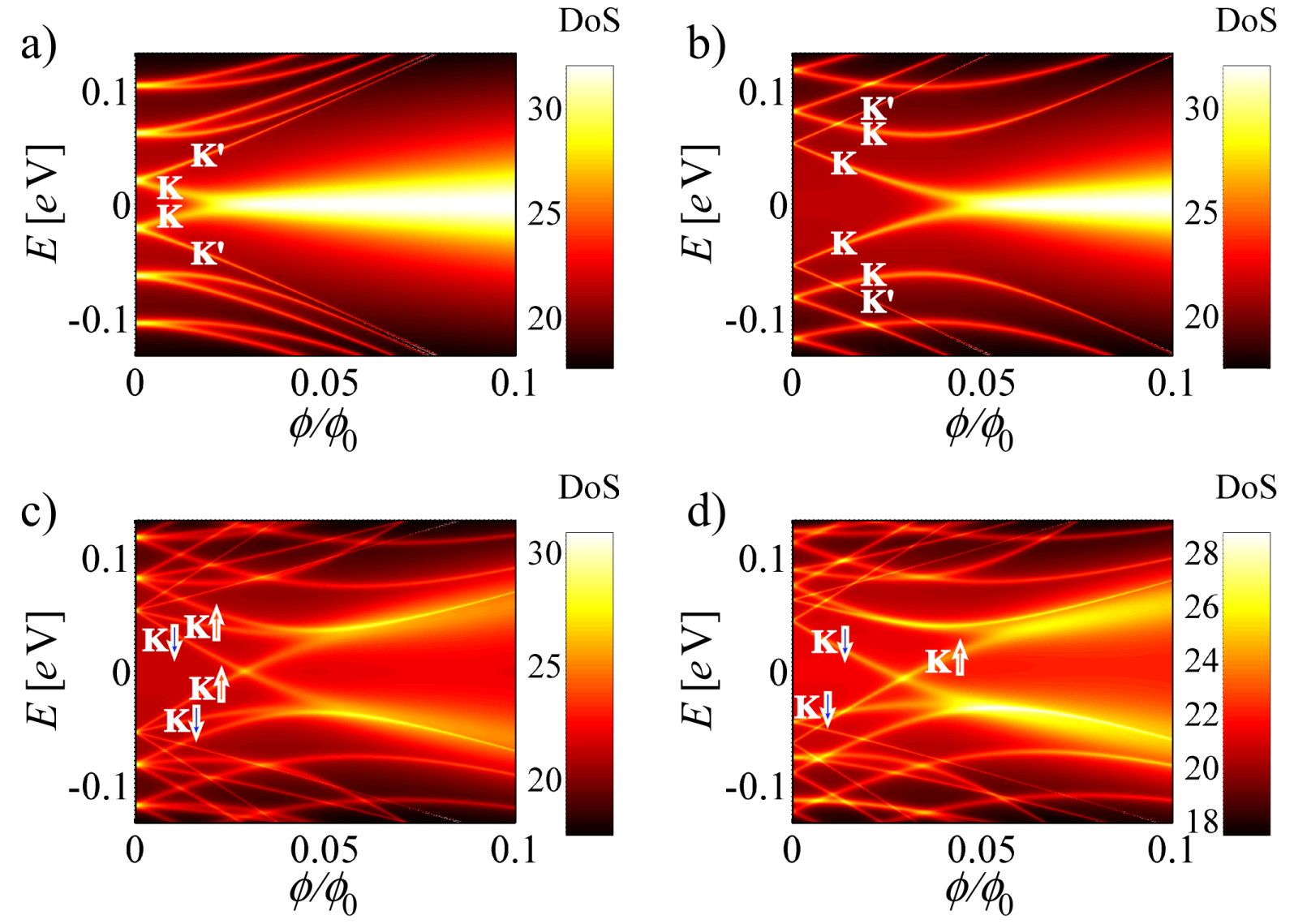}}
\caption{\label{fig:90_1}
Density of states of a (9,0) \CNT of a length of about 42 nm (100 unit cells) a) within a $\pi$-orbital tight-binding approximation, neglecting all other effects, b) with the inclusion of curvature effects as described in Sec.~\ref{ssec_model:curv}, c) and taking into account the Zeeman effect. Finally in d) spin-orbit coupling is additionally considered. The value of the \SOI parameter $\delta$ is set to $0.013$ and $\vect{\Sigma}_{{WB}} (E) \approx -\ii \,0.85 \eV$.
}
\end{figure}

\begin{figure}[h!]
\centerline{\includegraphics[width=\linewidth]{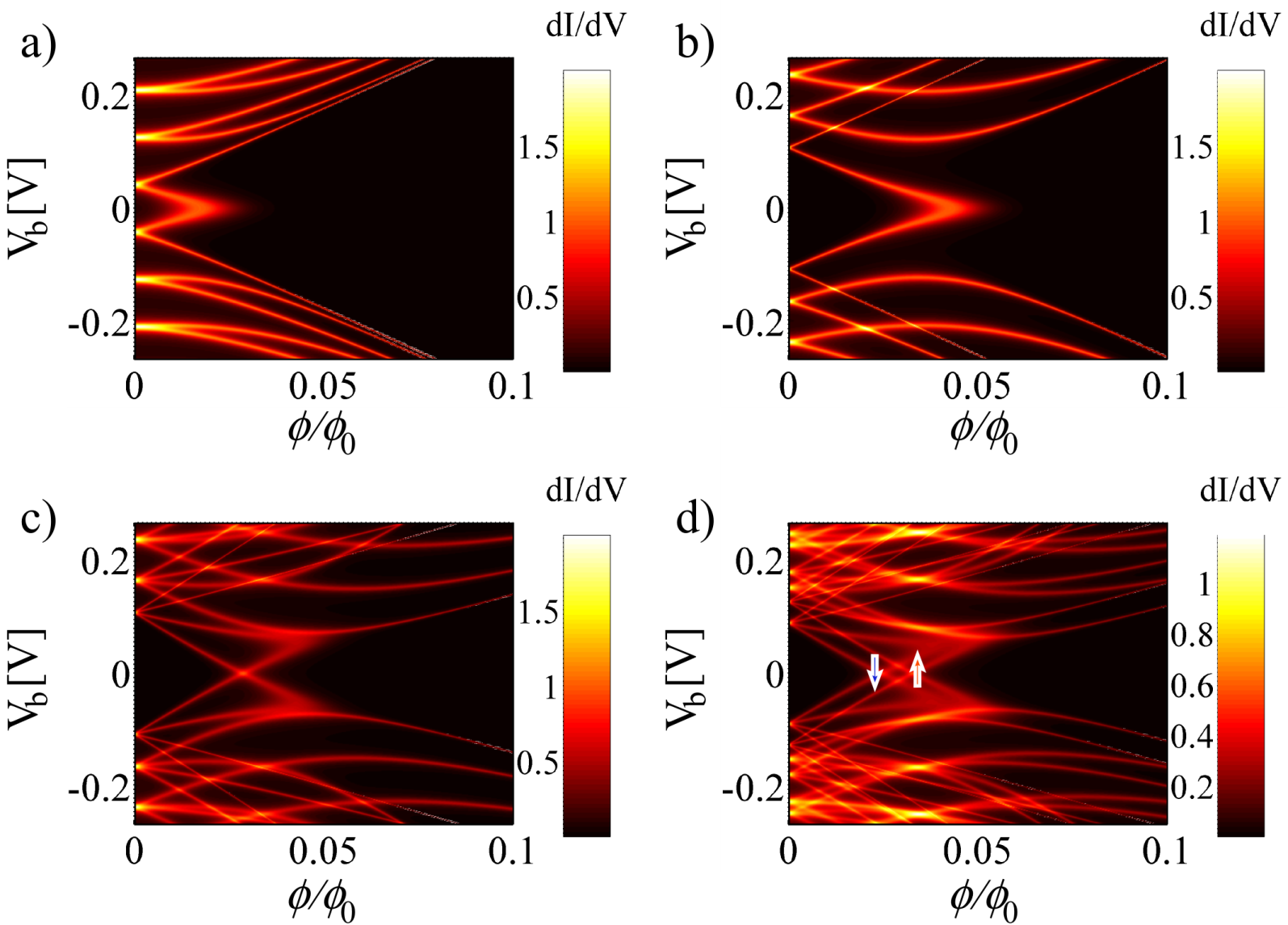}}
\caption{\label{fig:90_2}
$\textrm{d}I/\textrm{d}V$ characteristics in units of the conductance quantum $G_0$, for a (9,0) \CNT of a length of about 42 nm as in Fig.~\ref{fig:90_1} with the respective effects being successively included.
}
\end{figure}

With the help of the effective Hamiltonian and its eigenvalues given by Eq.~(\ref{eq:ana_final}) we can see that for the bare $\pi$-orbital model (Fig.~\ref{fig:90_1}a and Fig.~\ref{fig:90_2}a), there is a four-fold degeneracy at zero magnetic field: a spin- and a valley-degeneracy. The eigenvalues at $B=0$ correspond to the same allowed value of $\kappa_\perp$ (as we are dealing with a metallic \CNTns, specifically $\kappa_\perp=0$), and different allowed values of $\kappa_\parallel$ as determined by Eq.~(\ref{eq:BCzz_sol}). The valley-degeneracy is broken with the onset of the parallel magnetic field (Fig.~\ref{fig:90_1}a and Fig.~\ref{fig:90_2}a). As argued in Ref.~\onlinecite{MinotYSM04} and demonstrated analytically below, \cf Eq.~(\ref{eq:linearexp_orbmm}), this is the expected behavior if one assumes that i) the shift in energy due to the interaction of the orbital magnetic moment and the magnetic field is $\Delta E = -\vectGr{\mu}_{orb}\cdot\vect{B}$ and that ii) the orbital magnetic moment is reversed when moving from one Dirac cone or valley to the other. This implies that the states in the $\Kone$ Dirac cone will move towards the Fermi energy as their orbital magnetic moment is parallel to the magnetic field for conduction electrons and antiparallel for valence electrons. In the $\Ktwo$ cone the behavior will be inverted with states moving apart from the Fermi energy. This is indeed the evolution of the spectrum seen at low magnetic field. As the magnetic field is increased the evolution of the states becomes comparable in both valleys: all states move away from the Fermi energy, both in the valence and conduction band. We can understand this behavior in both regimes by looking at the limits of Eq.~(\ref{eq:BCzz_sol}). For small magnetic fluxes $\phi$, the components of the vector $\vectGr{\kappa}'$ can be written as $\kappa'_\perp\approx \frac{1}{\radio}\frac{\phi}{\phi_0}$ and $\kappa'_\parallel\approx \frac{\pi}{2L}(2n+1)+\Delta\kappa_\parallel$ with $n \in \mathbb{Z}$. In this limit Eq.~(\ref{eq:BCzz_sol}) yields $\Delta\kappa_\parallel= -\frac{\tau}{\radio}\frac{2}{\pi(2n+1)}\frac{\phi}{\phi_0}$. The spectrum given by Eq.~(\ref{eq:dr_general}) reads then to first order in $\phi$
\begin{equation}
\begin{split}
 E_\pm &= \pm \hbar \upsilon_F \sqrt{\lsb \frac{\pi}{2L}(2n+1)\rsb^2 -\frac{2\tau}{\radio L}\frac{\phi}{\phi_0}}\\
 &\approx \pm \hbar \upsilon_F \frac{\pi}{2L}(2n+1)\lsb 1 -\tau\frac{L}{\radio}\frac{4}{\pi^2 (2n+1)^2}\frac{\phi}{\phi_0}\rsb,
\end{split}\label{eq:linearexp_orbmm}
\end{equation}
that is, we have a linear evolution with the magnetic flux with slopes of different signs for the two independent Dirac cones. In other words we obtain the effective orbital moment at low fields being $\abso{\vectGr{\mu}_{orb}} = \frac{\radio e \upsilon_F}{\pi (2n+1)}$, which depends on the discrete value of $\kappa_\parallel$, and decreases as we move away from the band edges (corresponding to $\kappa_\parallel=0$). This is a signature of the finiteness of the tube, which could not be captured by the classical estimate of the orbital magnetic moment done in Ref.~\onlinecite{MinotYSM04}.

The finite values of $\kappa_\parallel$ described by Eq.(\ref{eq:BCzzkpar}) assure the presence of a gap in the absence of a magnectic flux, as observed in Fig.~\ref{fig:90_1}a and Fig.~\ref{fig:90_2}a.

At higher fields the evolution of the resonances with the magnetic flux is again linear. Now the allowed values of $\kappa'_\parallel$ remain nearly constant approaching asymptotically $\kappa'_\parallel \rightarrow \pm n\frac{\pi}{L}$ with $n >0 \in \mathbb{Z}$,
and the dominant contribution to the evolution of the states is given merely by $\kappa_\perp'$, that is, evolves linearly with a slope of $\frac{1}{\radio}$ as given by Eq.~(\ref{eq:kappaprimeTotal}). Moreover this slope is independent of the valley degree of freedom. At intermediate fields a crossover between the two different linear regimes appears.

As seen in Fig.~\ref{fig:90_1}b and Fig.~\ref{fig:90_2}b, the inclusion of the curvature modifies the spectrum of the \CNT widening the gap at zero magnetic flux. This can be understood by observing that $(\kappa'_\perp)^2$ is larger after the inclusion of $\Delta k_\perp^c$ (see Eq.~(\ref{eq:kappaprime})) and at the same time a larger value of $\abso{\kappa'_\perp}$ implies a larger $\abso{\kappa'_\parallel}$ (see Fig.~\ref{fig:allowedk}). The states are still spin-degenerate. We also observe that the states now cross as they evolve with the magnetic flux. The crossings are possible because of the curvature-induced shift $\Delta k_\perp^c$ of the perpendicular momentum, which is valley dependent. We have therefore two different evolutions for $\kappa_\perp'$ for the $\Kone$ and $\Ktwo$ cones, which allow crossings of states of different valleys.

In Fig.~\ref{fig:90_1}c and Fig.~\ref{fig:90_2}c we introduce the Zeeman effect in this picture. The spin-degeneracy breaks at finite fields, and an additional evolution, opposite for spins up and down, competes with the evolution due to the orbital magnetic moment.

Finally, \SOI introduces a splitting of the spin species already at zero magnetic field (Fig.~\ref{fig:90_1}d and Fig.~\ref{fig:90_2}d). Now $\kappa'_\perp$ is shifted towards negative values for spin up and towards positive values for spin down.  A detailed analysis of the momentum shift caused by the different effects combined with the diagram of solutions for $\lcb \kappa'_\perp,\kappa'_\parallel\rcb$ (Fig.~\ref{fig:allowedk}), allows us to conclude that now the degeneracy at zero field is between a $\Kone\!\downarrow$ and a $\Ktwo\!\uparrow$ subband and between a $\Kone\!\uparrow$ and a $\Ktwo\!\downarrow$ subband. This degeneracy is broken at finite magnetic fields. Including the Zeeman effect introduces a different evolution with the magnetic field for the two spin species. This leads to the breaking of the electron-hole symmetry in the spectrum of the \CNT
as it can be seen in Fig.~\ref{fig:90_1}d. In the corresponding $\textrm{d}I/\textrm{d}V$ characteristics we recover a symmetric picture, as we are assuming a symmetric voltage drop across the junction. The chemical potential of the right and left lead is set at $E_F \pm \frac{\eV_b}{2}$ with $V_b$ being the bias voltage. In this case at zero temperature the Landauer formalism allows us to write the differential conductance as
\begin{equation}
  \frac{dI}{dV}= \frac{2 e^2}{h} \frac{T\lrb E_F+\frac{\eV_b}{2}\rrb +T\lrb E_F-\frac{\eV_b}{2}\rrb}{2},
\end{equation}
which is symmetric in the bias independent of the transmission $T$.

The Zeeman splitting leads as well to the breaking of the quantum flux periodicity in the magnetic field, which has been checked numerically (not visible in Fig.~\ref{fig:90_1} and Fig.~\ref{fig:90_2} where a smaller range of the magnetic field has been chosen for clarity).

\section{Localization in the magnetic field}\label{sec_numerical_loc}

\begin{figure}[h!]
\centerline{\includegraphics[width=\linewidth]{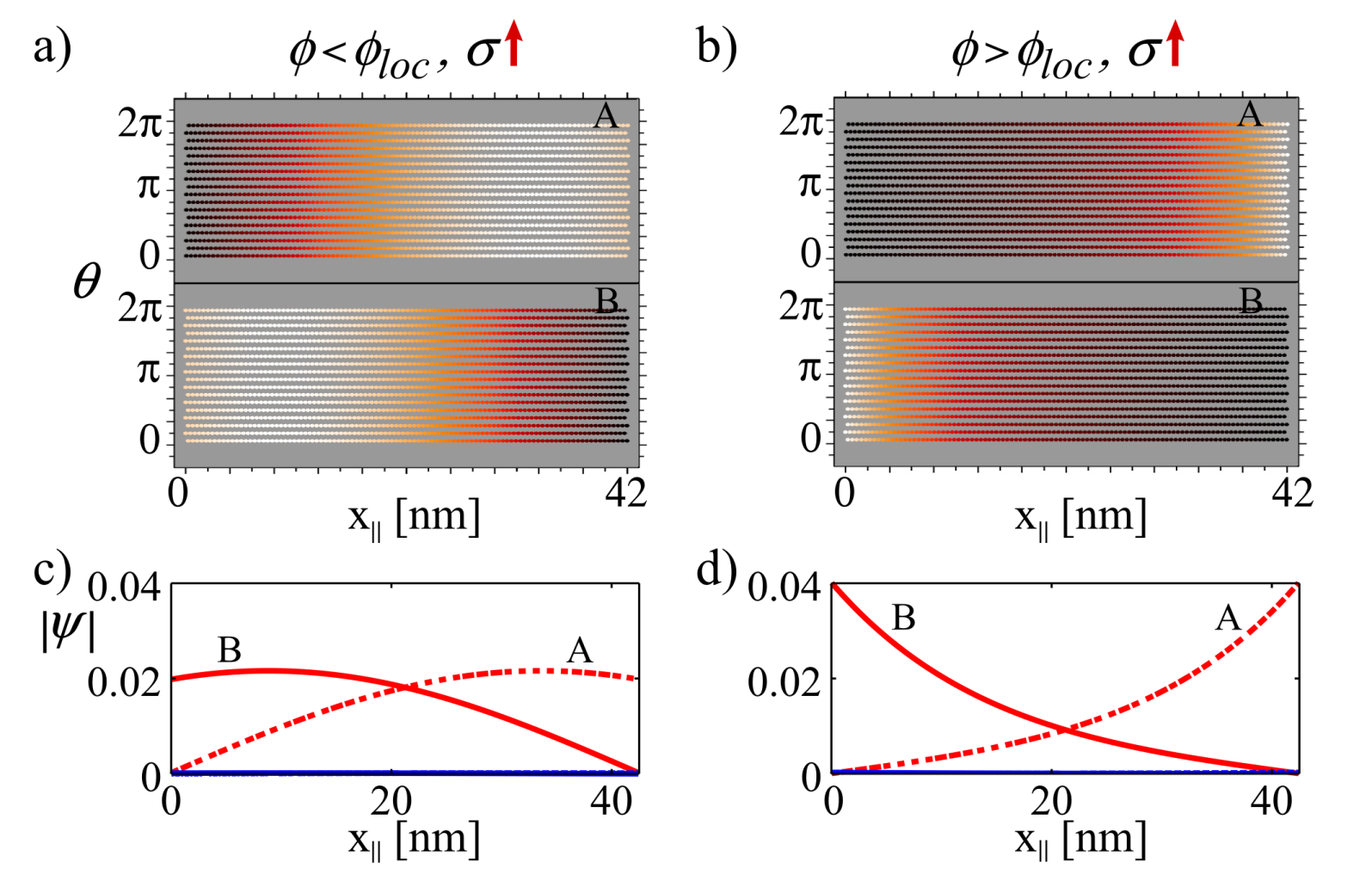}}
\caption{\label{fig:LocalizationOnEdges}
Spatially resolved wave function (obtained numerically) of a (9,0) CNT with 100 unit cells and \SOI corresponding to the valence eigenstate getting localized as a spin-up state (see Fig.~\ref{fig:90_1}).  In the top figures, dots represent the atomic positions and their color corresponds to the amplitude of the wave function at the site (going from zero to its maximal value). The two sublattices $A$ and $B$ are plotted separately. The corresponding magnetic fluxes are a) $\phi=0.02\phi_0$ and b) $\phi=0.05\phi_0$. The decay of the wave function amplitude for these two flux values is shown in c) and d) respectively, where sublattice $A$ is plotted with a dashed line and a solid line is used for the results corresponding to sublattice $B$. Red marks the spin-up component of these states while blue is used for spin-down, which is negligible here.
}
\end{figure}

In Fig.~\ref{fig:90_1} and Fig.~\ref{fig:90_2} we observe a remarkable feature. The density of states acquires a pronounced peak at zero energy above a threshold value of the magnetic field.
Once the Zeeman effect is considered, this peak is split into two, appearing now at finite energies.
At the same time we observe that the differential conductance for these states is strongly suppressed.
This feature reflects the onset of the localized solutions~\cite{SasakiMSK05,MarganskadVJSG11} described by Eq.~(\ref{eq:BCzz_im}). States with imaginary $\kappa'_\parallel$ occur in the vicinity of the Fermi point $\Kone$ at $\kappa'_\perp=1/L$, and for $\Ktwo$ instead at  $\kappa'_\perp=-1/L$. The threshold value in the magnetic field is therefore given by
\begin{equation}
     \frac{\phi_{loc}}{\phi_0} = \radio \lrb \tau\frac{1}{L}-\kappa_\perp - \Delta k_\perp^c - \sigma\Delta k_\perp^{SO}\rrb,
\end{equation}
where we have made use of Eq.~(\ref{eq:kappaprimeTotal}).
That is, the onset for the imaginary solution in the $\Kone$ cone is around $\radio/L$ with a shift towards larger fields depending on the chirality (being maximal for zigzag \CNTsns), and a further shift depending on the \SOI parameter $\delta$ towards larger fields for spin up and towards lower fields for spin down. Analogously for $\Ktwo$.

These eigenstates gradually get localized at the edges of the \CNT and we can see in Fig.~\ref{fig:LocalizationOnEdges} the onset of this localization for the spin-up state of the valence band which localizes above the Fermi energy. Fig.~\ref{fig:LocalizationOnEdges}a corresponds to a magnetic flux below the threshold value $\phi_{loc}$ and the magnetic flux in Fig.~\ref{fig:LocalizationOnEdges}b is already above this value, which for this \CNT is $\phi_{loc}\approx 0.043 \phi_0$ (for $\tau=1$ and $\sigma=1$). We therefore see in the decay of the wave function amplitude (Fig.~\ref{fig:LocalizationOnEdges}c and Fig.~\ref{fig:LocalizationOnEdges}d) its change towards a localization at the edges with an exponential decay for the larger magnetic field value as it corresponds to an imaginary momentum.
The $A$ and $B$ components still overlap for this magnetic field, so that the corresponding conductance is reduced but not suppressed yet. In both cases they are eigenstates of the angular momentum in the perpendicular direction, corresponding to rotations around the tube axis, so that the amplitude is constant over the angular variable $\theta$. For $\phi\gtrsim 0$ the $\Kone$-$\Ktwo$ degeneracy
is broken, and thus this angular distribution is expected.

The eigenstates do not have a well-defined spin because of the \SOIns. However we can refer here to a ``spin-up'' state because the amplitude of the spin down component is several orders of magnitude smaller than that of the up component (see panels c and d of Fig.~\ref{fig:LocalizationOnEdges}). For the corresponding valence eigenstate localizing as a spin down we obtain the same picture with just a different value for the onset of the localization ($\phi_{loc}\approx 0.029 \phi_0$ for $\tau=1$ and $\sigma=-1$), so that for the magnetic flux of $\phi=0.05\phi_0$, the localization at the edges is stronger.

Many nanotubes (most notably zigzag \CNTsns) present ``native`` localized states, present at their edges even in the absence of the magnetic field.~\cite{FujitaWNK96,NakadaFDD96,WakabayashiFAS99,KimOHL99,OkadaO03} They correspond to states from higher energy subbands, with $\kappa_\perp = \pm\frac{1}{\radio}$, where $\kappa'_\perp$ fulfills the localization condition naturally. The native localized states are not seen in these figures because of the rather strong coupling to the leads, but are recovered if we reduce it.

\begin{figure}[h!]
\centerline{\includegraphics[width=\linewidth]{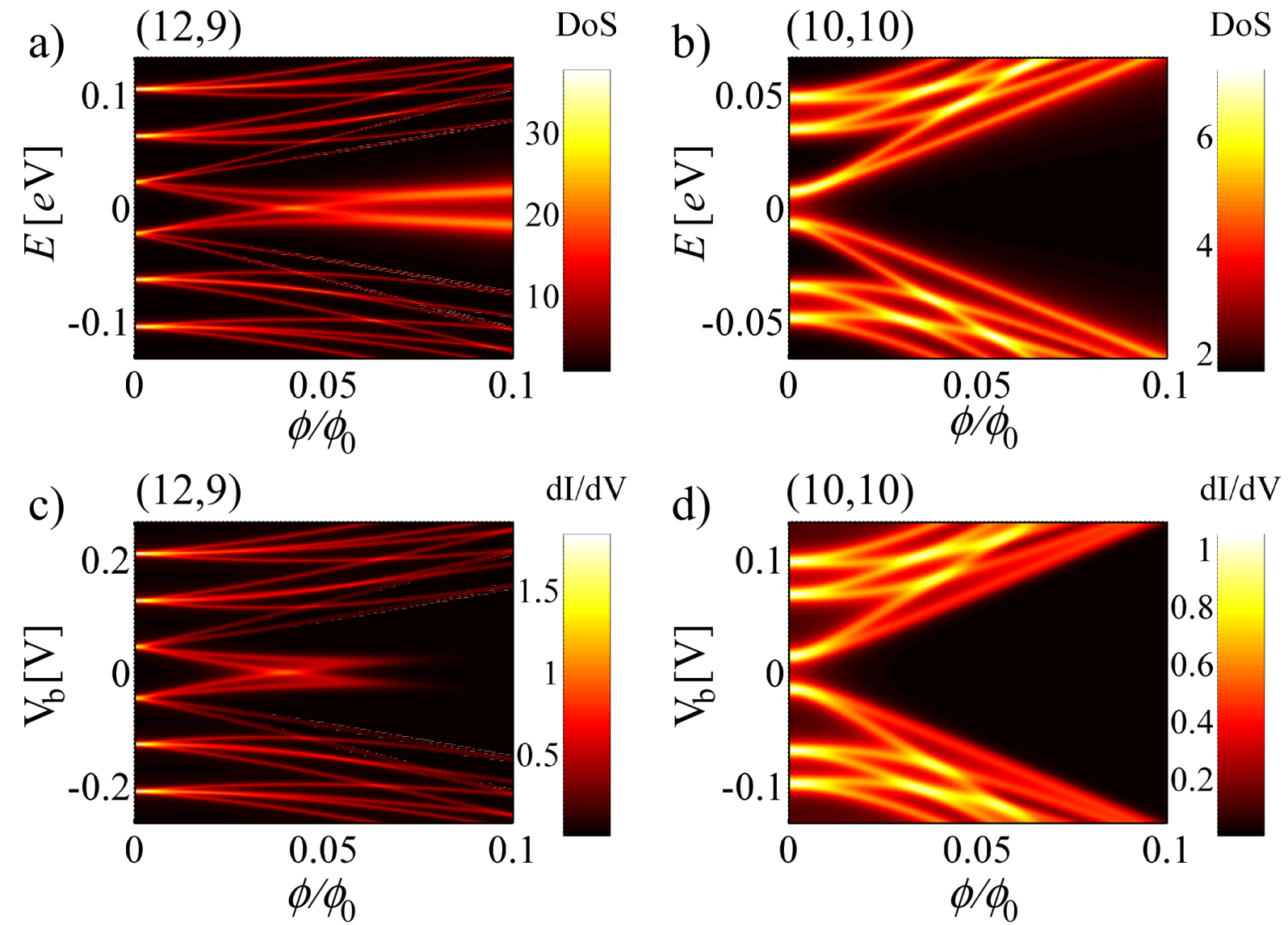}}
\caption{\label{fig:otherChiralities}
a) Density of states of a chiral (12,9) \CNT ($\eta=25^\circ$) with a length of 16 unit cells (about 42 nm) taking into account curvature effects, the spin-orbit coupling and the Zeeman effect. b) Density of states of an armchair \CNT (10,10) of 172 unit cells (about the same length of 42 nm) and with the same approximations as in (a). c) and d) are the $\textrm{d}I/\textrm{d}V$ characteristics in units of $G_0$ corresponding respectively to a) and b).  The value of the \SOI $\delta$ is set to $0.0028$. Notice the absence of localization in the armchair case.
}
\end{figure}

This localization induced in the magnetic field is also observed in chiral nanotubes as mentioned before in Sec.~\ref{sec_model:dispersion_BC}. We can see in Fig.~\ref{fig:otherChiralities}a,c the case of a (12,9) chiral \CNTns, which has a chiral angle already quite close to the one of armchair nanotubes, and for which the localization is clearly visible. For armchair nanotubes on the other hand (Fig.~\ref{fig:otherChiralities}b,d) this localization is not possible, as the boundary conditions do not allow for imaginary solutions of the $\vectGr{\kappa}$ vector.

\begin{figure}[h!]
\centerline{\includegraphics[width=\linewidth]{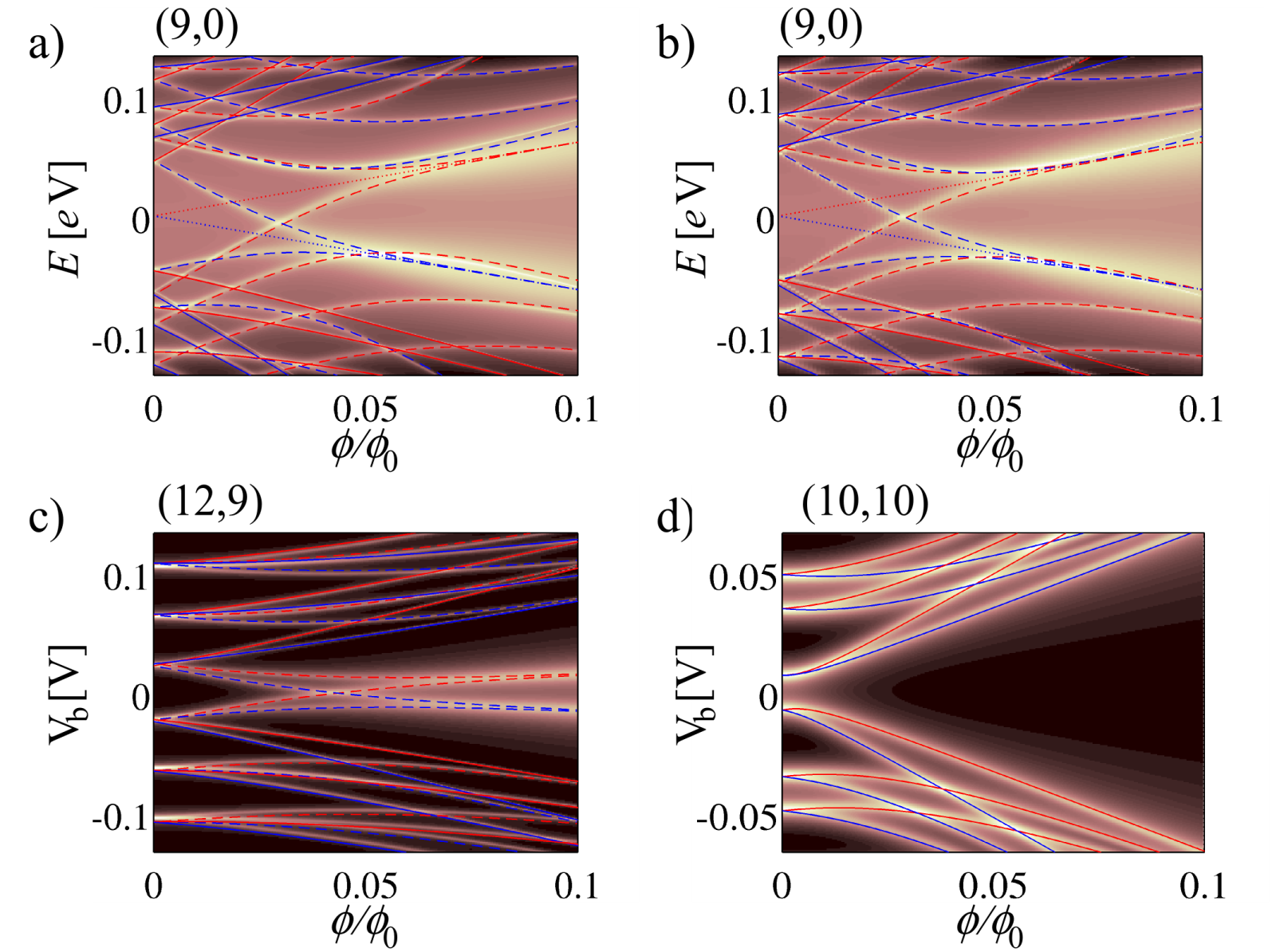}}
\caption{\label{fig:comparisonAnalytics}
Comparison of numerical and analytical results for the projected density of states in different \CNTs of approximately the same length (about 42 nm). The dashed lines in the analytical calculation correspond to states in the $\Kone$ cone and the solid lines correspond to $\Ktwo$ states. Spin up are represented in red, and spin down in blue. All the lines shown correspond to the band with $\kappa_\perp=0$, except for the states appearing in the zigzag \CNT at the Fermi energy at $\phi=0$, which are the edge states corresponding to neighboring bands (the next allowed $\kappa_\perp$ values, $\kappa_\perp=\pm\frac{1}{\radio}$) and for which dotted lines are used. a) A (9,0) \CNT as in Fig.~\ref{fig:90_1}d.  b) The same (9,0) \CNT but with a more realistic value of $\delta$, as in the calculations shown in Fig.~\ref{fig:otherChiralities}. c) A (12,9) \CNT as in Fig.~\ref{fig:otherChiralities}a. d) A (10,10) \CNT as in Fig.~\ref{fig:otherChiralities}b.
}
\end{figure}

The very good agreement between numerical and analytical results is clearly seen from Fig.~\ref{fig:comparisonAnalytics}. It makes evident that the spin-flipping part of the Hamiltonian has a very small contribution so that we are justified to assign a specific spin to the transport features. In this way we can see that it is possible to have a control of the valley and spin degrees of freedom in \CNTs due to the presence of \SOIns. This feature will thus be exploited in the next section.

\section{Generation of spin-polarized currents}\label{sec_numerical_pol}
%
\begin{figure}[h!]
\centerline{\includegraphics[width=\linewidth]{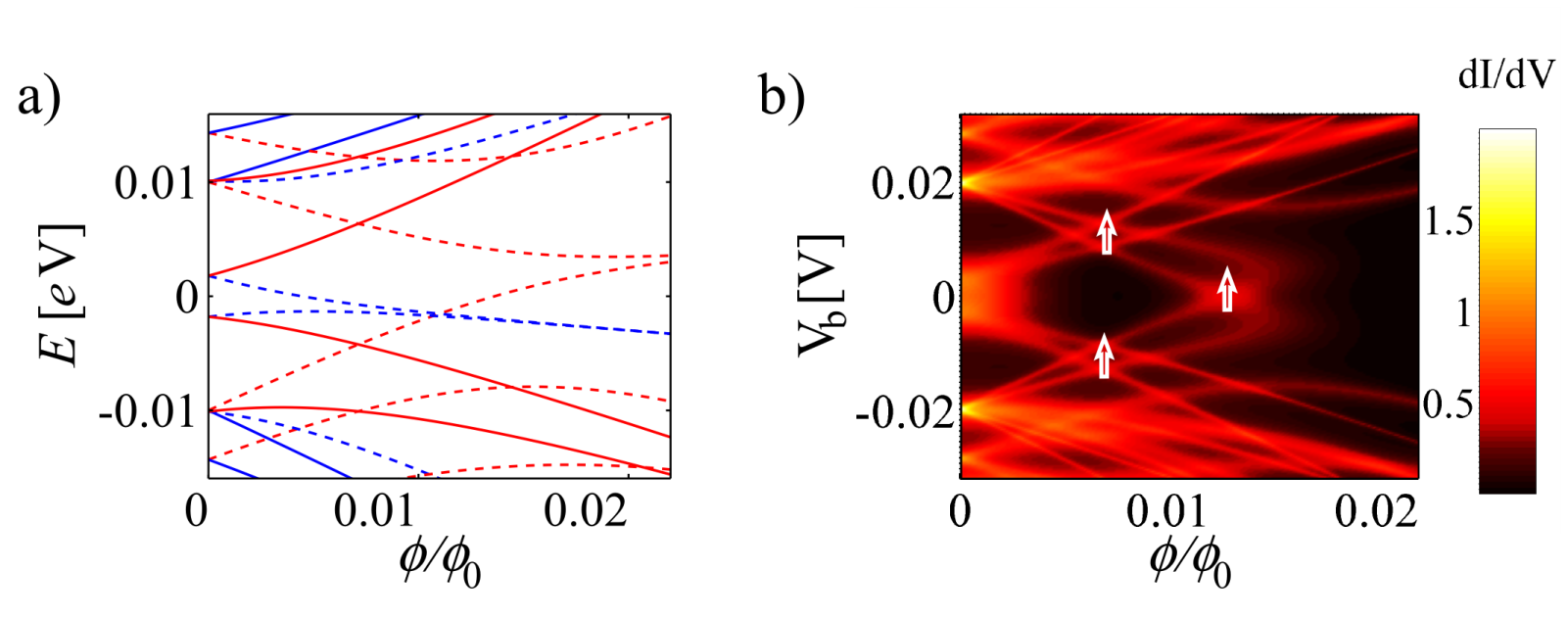}}
\caption{\label{fig:polarization}
a) Spin-resolved energy spectrum for the \CNT (12,9) with a length of 100 unit cells and a \SOI parameter $\delta$ of 0.013. The colors correspond to those in Fig.~\ref{fig:comparisonAnalytics}.
b) Corresponding differential conductance, where the spectrum lines have been symmetrized due to the considered symmetrical voltage drop.
}
\end{figure}

\begin{figure}[h!]
\centerline{\includegraphics[width=\linewidth]{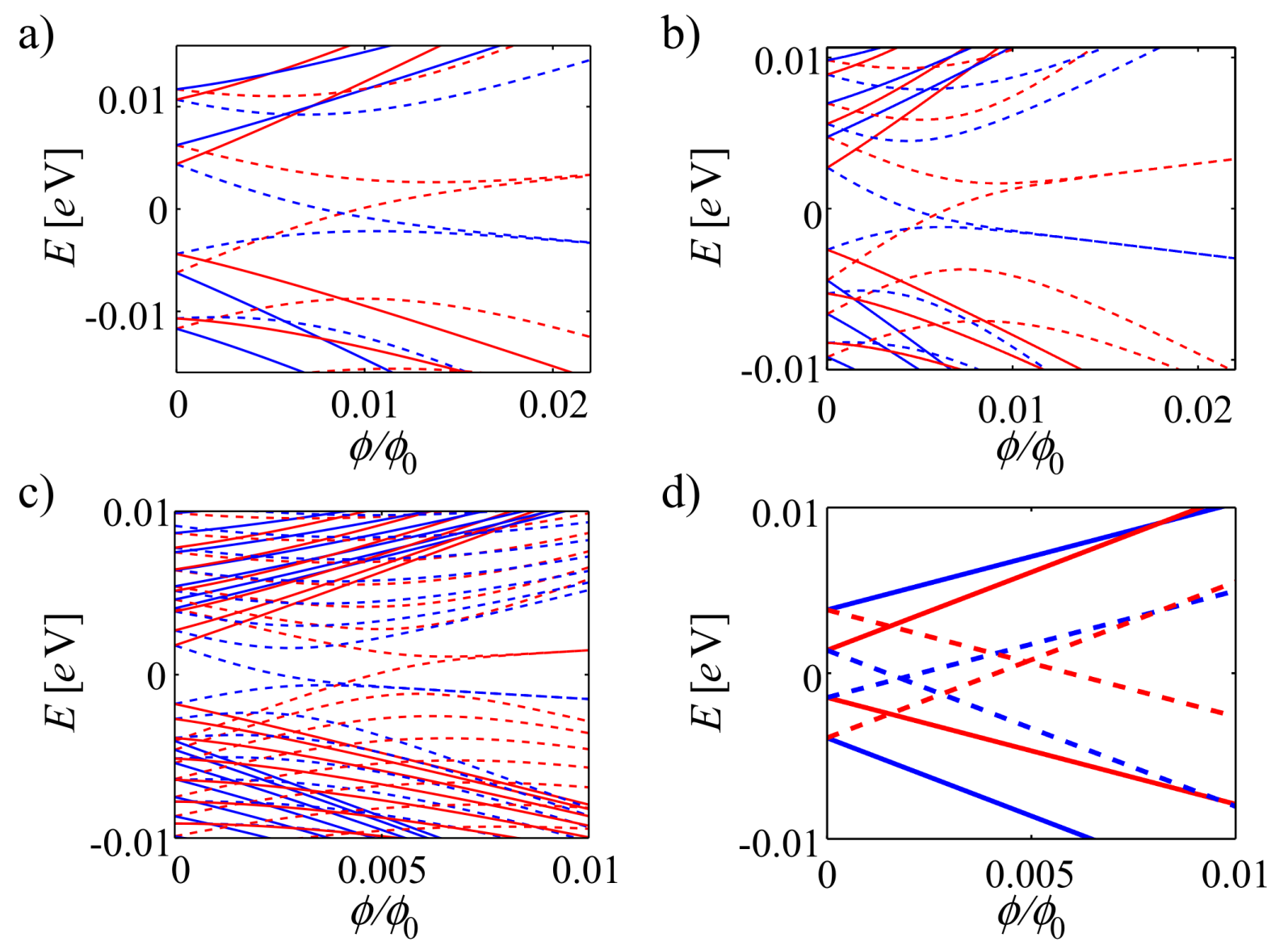}}
\caption{\label{fig:129_Lengths}
Spin-resolved energy spectrum for (12,9) \CNTs of several lengths with a \SOI characterized by the parameter $\delta=0.0028$. a) A hundred unit cell long \CNTns, that is, about 259 nm long. b) A \CNT with 200 unit cells, and thus a length of about 518 nm. c) \CNT with a length of about 1295 nm (500 unit cells). d) Evolution with the magnetic field of the band edges for an infinite \CNTns.
The colors correspond to those in Fig.~\ref{fig:comparisonAnalytics}.}
\end{figure}

The presence of localized states of definite spin, suggests the possibility of generating polarized spin currents, with in addition a definite isospin, by exploiting the localization phenomenon. For example, if a spin up current is desired, one should look for \CNTs such that the localization onset of the $\Kone\!\downarrow$ states occurs already at negative magnetic fields. This condition would lead to a region around the Fermi energy of highly spin-up polarized current for positive magnetic fields. Correspondingly spin-down polarized current characterizes negative fields. This possibility is demonstrated in Fig.~\ref{fig:polarization}a with $\phi_{loc}(\Kone\!\downarrow)=-0.001\phi_0$, calculated for clarity with a large value of $\delta$. In a wide energy window around the Fermi energy all transport features are characterized by a spin up, with the only exception of the states $\Kone\!\downarrow$ which are governed by wave functions decaying already exponentially from the edge states as seen in Fig.~\ref{fig:polarization}b. We can achieve this state also with a realistic $\delta$ by tuning the other parameters influencing $\phi_{loc}$, as the length of the \CNTns.
An example of a \CNT also fulfilling these conditions is for instance a (38,37) \CNT with a length of 100 unit cells. Its chirality is very close to that of armchair \CNTsns, which minimizes the shift $\Delta k_{\perp}^c$, and together with its long length allows for a localization onset of $\Kone\!\downarrow$ at negative fields. A greater length is of advantage because for larger \CNTs the onset of the localization is moved towards smaller magnetic fields, and thus corresponds to a smaller Zeeman shift. We can see this effect in Fig.~\ref{fig:129_Lengths}, where the spectra for (12,9) chiral \CNTs with different lengths are shown. The features of the resonances evolve more and more towards the results expected for the infinite \CNTs (Fig.~\ref{fig:129_Lengths}d), but showing the localization induced by the magnetic field at finite lengths.

\begin{figure}[h!]
\centerline{\includegraphics[width=\linewidth]{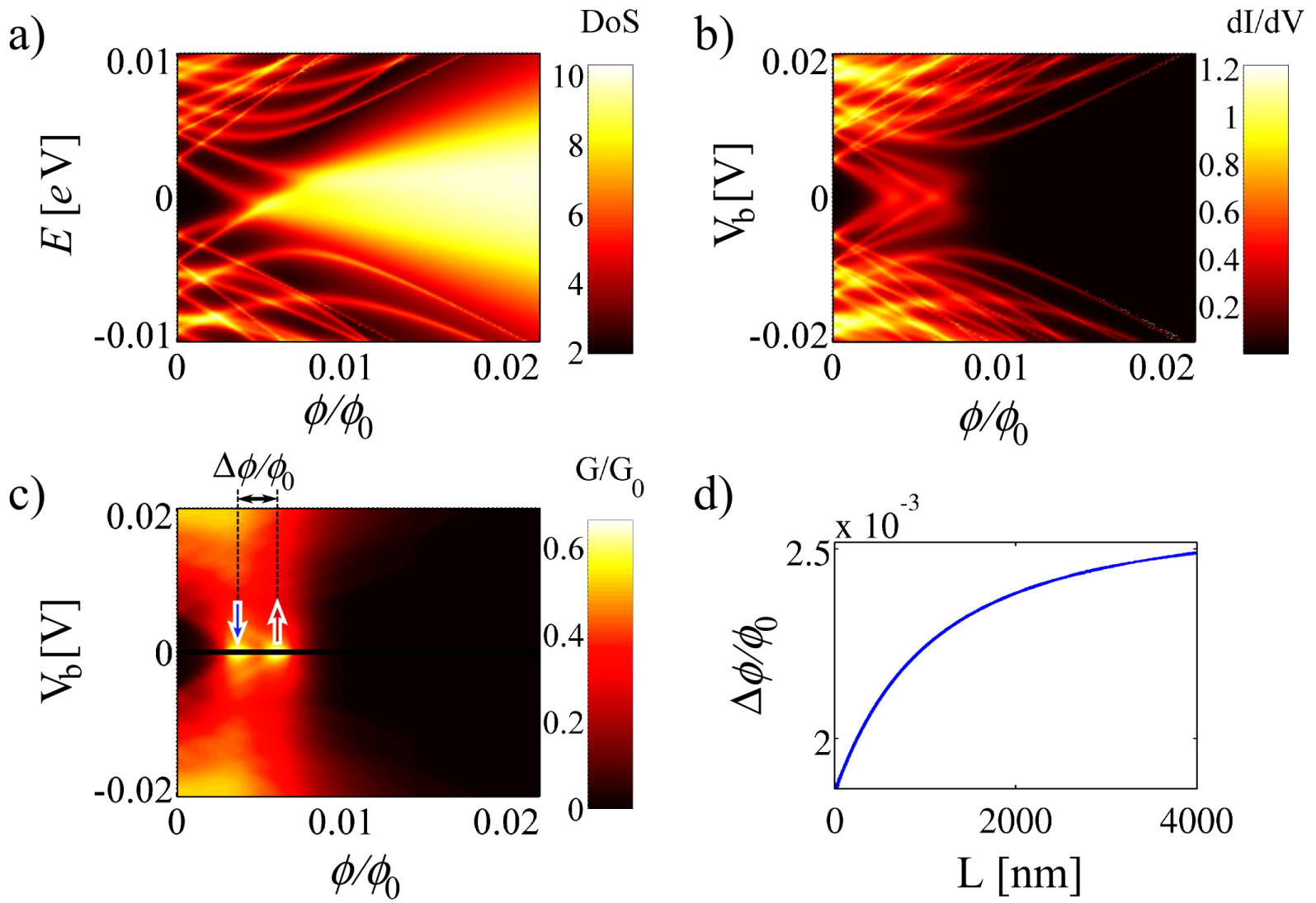}}
\caption{\label{fig:129_200uc}
a) Evolution of the spectrum in the magnetic field and b) differential conductance as well as c) integrated conductance for a (12,9) \CNT of 200 unit cells. The points around $V_b=0$ have been suppressed in order to avoid a division by zero. d) Growth with the \CNT length of the separation in magnetic flux of the conductance peaks around the Fermi energy.
}
\end{figure}

Another kind of polarization shows up in these systems as a result of the combination of the \SOI and the localized states in the magnetic field. Without the \SOI we have observed in Fig.~\ref{fig:90_1}c that the localizing states cross the Fermi energy in the presence of the Zeeman effect at the same field intensity for the $\Kone\!\downarrow$ state of the conduction band and the $\Kone\!\uparrow$ state of the valence band. At this crossing point we have therefore a mixture of spin-up and spin-down states. The inclusion of \SOI breaks the electron-hole symmetry of this picture and this crossing point is split into two, as the $\Kone\!\downarrow$ state lying closer to the Fermi energy crosses it at a lower magnetic field than the $\Kone\!\uparrow$ state. In a small bias window we thus obtain a spin-down polarized current swapping into a spin-up polarized current with increasing magnetic field. These two current peaks can be resolved in the magnetic flux as seen in Fig.~\ref{fig:129_200uc}. For negative magnetic fields both the spin and the isospin of these states will be reversed. Fig.~\ref{fig:129_200uc} shows a \CNT of 200 unit cells as the one described in Fig.~\ref{fig:129_Lengths}b. The distance between these peaks of polarized current is set by the \SOI parameter as well as the characteristics of the tube, such as length, radius and chirality. This separation of the peaks in the magnetic field will increase with the \CNT length as seen in Fig.~\ref{fig:129_200uc}d.

\section{Conclusions}\label{sec_concl}

We have calculated the energy spectrum and transport properties of finite carbon nanotubes under a parallel magnetic field, including spin-orbit interactions. The spin-orbit coupling was treated following the approach of Ando,~\cite{Ando00} but the spin quantization axis has been chosen parallel to the magnetic field to conveniently incorporate its effects. The mixing of the $\pi$ and $\sigma$ orbitals, due to the curvature of the nanotube, is automatically taken into account by this approach.

We derived an analytical formula for the energy spectrum of these systems, which provides a very good matching to the numerical calculations in its limits of validity, \ie for large enough radii to justify the approximations made. For radii of about 0.4 nm the agreement is already over 90\%.
This analytical formula can be applied to finite-size systems if the proper boundary conditions are imposed. The very good results of this approach allow us to use it as a powerful tool to gain insight into the numerical results, as we can identify the composition of the states in the energy spectrum without the highly costly numerical calculation of the eigenstates. We can furthermore vary the length of the central system at no computational cost.

The analysis of the transport properties (\eg the $\textrm{d}I/\textrm{d}V$ characteristics) shows a mechanism of localization in chiral and zigzag nanotubes in a parallel magnetic field. The magnetic field modifies the boundary conditions, causing hitherto extended states to localize near the
ends of the tubes. This localization is gradual and initially the states involved are still conducting. But at a critical
value of the magnetic field, which depends on the nanotube chirality and length, the localization is
complete and the transport is then suppressed.

Furthermore the \SOI together with the Zeeman effect introduces a spin-splitting that breaks the electron-hole symmetry and allows the separation of polarized states. By exploiting this effect we achieve spin polarized currents across the system. The magnetic field can thus be used as a tool for controlling the spin and isospin of the current through the \CNTns.

\begin{acknowledgments}
This work was partially funded by the \textit{Deutsche Forschungsgemeinschaft} under the program GRK1570.
\end{acknowledgments}

\appendix
\section{Modification of the hopping integrals by the \SOI}\label{sec_appA}

In the lowest order perturbation theory the state $\ket{z_j \uparrow_y}$ is given by
\begin{equation}
  \ket{z_j \uparrow_y} \approx \rket{z_j \uparrow_y} +
\sum_{\substack{\alpha = x,y,z\\ s =\uparrow_y,\downarrow_y \\\alpha s \neq z \uparrow_y}}
  \rket{\alpha_j s_j}\frac{\rbra{\alpha_j s_j} \Ham_{so} \rket{z_j \uparrow_y}}{\varepsilon_{2p}^\pi - \varepsilon_{2p}^{\alpha}}.
\end{equation}
The kets $\rket{\uparrow_y}$ and $\rket{\downarrow_y}$ are the two eigenstates of the Pauli's matrix $\sigma_{y}$, the normalized eigenspinors in the $y_j$ direction which coincides with the global $Y$. They can be expressed as functions of the eigenspinors in the $z_j$ direction, the original axis of spin quantization, as:
\begin{equation}
\begin{split}
  \rket{\uparrow_y}   &=  \frac{\ee^{\ii\theta_j/2}}{\sqrt{2}} \lcb \rket{\uparrow_j} + \ii\rket{\downarrow_j}\rcb,\\
  \rket{\downarrow_y} &=  \frac{\ee^{-\ii\theta_j/2}}{\sqrt{2}} \lcb \rket{\uparrow_j} - \ii\rket{\downarrow_j}\rcb.
\end{split}
\end{equation}

The action of Pauli's spin matrices $\sigma_x$ and $\sigma_z$ on these spinors produces a spin flip whereas $\sigma_y$ conserves the spin.
Because of the orthogonality of the spin eigenstates and the symmetry of the orbital parts of the integrals, we can see that most terms will vanish, leaving only two contributions:
\begin{equation}
\begin{split}
  \ket{z_j \uparrow_y} \approx \rket{z_j \uparrow_y} &+ \rket{x_j \uparrow_y}\frac{\rbra{x_j \uparrow_y} \Ham_{so}^y \rket{z_j \uparrow_y}}{\varepsilon_{\pi\sigma}}\\
 &+ \rket{y_j \downarrow_y}\frac{\rbra{y_j \downarrow_y} \Ham_{so}^x \rket{z_j \uparrow_y}}{\varepsilon_{\pi\sigma}},
\end{split}
\end{equation}
where $\Ham_{so}^\alpha= \frac{\hbar}{4 m^2 c^2} \lrb \nabla V \times \vect{p} \rrb_\alpha \sigma_\alpha$ and  $\varepsilon_{\pi\sigma}=\varepsilon_{2p}^\pi - \varepsilon_{2p}^\sigma > 0$ is the energy difference between the $p_z$ and $sp^2$ orbitals, which are building respectively the $\pi$ and $\sigma$ molecular orbitals.  It can already be predicted by symmetry that the orbital contributions of the integrals are equal, so that we can define
\begin{equation}
 \varepsilon_{\pi\sigma}\delta =  -\ii\rbra{x_j \downarrow_y} \Ham_{so}^y \rket{z_j \uparrow_y} =
 \ee^{-\ii\theta_j} \rbra{y_j \downarrow_y} \Ham_{so}^x \rket{z_j \uparrow_y},
\end{equation}
in order to write the modified $\pi$-states as
\begin{equation}
  \ket{z_j \uparrow_y}  \approx  \rket{z_j \uparrow_y} + \ii\delta\rket{x_j \uparrow_y} + \delta e^{\ii\theta_j} \rket{y_j \downarrow_y},
\label{eq:Andos_zju}
\end{equation}
and similarly for spin down
\begin{equation}
  \ket{z_j \downarrow_y} \approx  \rket{z_j \downarrow_y} - \ii\delta\rket{x_j \downarrow_y} - \delta e^{-\ii\theta_j} \rket{y_j \uparrow_y}.
\label{eq:Andos_zjd}
\end{equation}
We have arrived at an expression for the perturbed orbitals in terms of a global spin direction. The parameter $\delta$ contains then the measure of the strength of the intra-atomic spin-orbit coupling.

We proceed now to calculate the modified hopping parameters between these states, which, after carrying out the spin part of the integrals and discarding higher order terms in $\delta$, are given by
\begin{eqnarray}
 \bra{z_i\uparrow_y}\, \Ham\, \ket{z_j\uparrow_y} &=& \rbra{z_i}\Ham\rket{z_j} \nonumber\\
&&+ \ii\delta\;\bigl\{ \rbra{z_i}\Ham\rket{x_j} - \rbra{x_i}\Ham\rket{z_j}\bigr\}, \nonumber\\
 \bra{z_i\downarrow_y}\, \Ham\, \ket{z_j\downarrow_y} &=& \rbra{z_i}\Ham\rket{z_j} \nonumber\\
&&- \ii\delta\;\bigl\{ \rbra{z_i}\Ham\rket{x_j} - \rbra{x_i}\Ham\rket{z_j}\bigr\},\nonumber\\
 \bra{z_i\uparrow_y}\, \Ham\, \ket{z_j\downarrow_y} &=&
     \,\delta\;\ee^{-\ii\theta_i}\rbra{y_i}\Ham\rket{z_j} \nonumber\\
&&- \delta\;\ee^{-\ii\theta_j} \rbra{z_i}\Ham\rket{y_j},\nonumber\\
 \bra{z_i\downarrow_y}\, \Ham\, \ket{z_j\uparrow_y} &=&
  \,\delta\;\ee^{\ii\theta_j} \rbra{z_i}\Ham\rket{y_j} \nonumber\\
&&- \delta\;\ee^{\ii\theta_i}\rbra{y_i}\Ham\rket{z_j}.
\label{eq:NewHop1}
\end{eqnarray}
The last two hopping integrals are the spin flipping terms, which the \SOI make possible. They are a consequence of the spin mixing appearing in Eq.~(\ref{eq:Andos_zju}) and Eq.~(\ref{eq:Andos_zjd}), where we see that the states including the \SOI do not have a well-defined spin, but there is some admixture of the opposite spin species.

In order to calculate the orbital part of the transfer integrals in Eq.~(\ref{eq:NewHop1}) we observe that
the hopping $\rbra{\alpha_i}\Ham\rket{\alpha_j}$ between the neighbor orbitals $\rket{\alpha_i}$ and $\rket{\alpha_j}$, $\alpha=x,y,z$, can be written in terms of its normal and tangential components in the graphene plane building the nanotube, as we have already seen in Sec.~\ref{ssec_model:curv}, for the case of $\rket{z_j}$ orbitals. Explicitly,
\begin{equation}
\begin{split}
\rbra{\alpha_i}\,\Ham\,\rket{\alpha_j} &=
V_{pp}^\pi\; \vect{n}(\alpha_i)\cdot\vect{n}(\alpha_j)\\
&+ (V_{pp}^\sigma - V_{pp}^\pi) \frac{(\vect{n}(\alpha_i)\cdot\vect{R}_{ji}) (\vect{n}(\alpha_j)\cdot\vect{R}_{ji})}{\abso{R_{ji}}^2}\quad \quad
\end{split}
\label{eq:oti2}
\end{equation}
for generic $p_\alpha$ orbitals.

Thus we find
\begin{eqnarray}
 &&\rbra{z_i}\Ham\rket{x_j} = -\rbra{x_i}\Ham\rket{z_j} = V_{pp}^\pi \sin{\lrb\theta_i-\theta_j\rrb} \nonumber\\
 &&\quad\quad-(V_{pp}^\sigma - V_{pp}^\pi) \frac{\radio^2}{\aC^2} \sin{\lrb\theta_i-\theta_j\rrb} \lsb \cos{\lrb\theta_i-\theta_j\rrb}-1\rsb,\nonumber\\
 &&\rbra{z_i}\Ham\rket{y_j} = -\rbra{y_i}\Ham\rket{z_j} \nonumber\\
          &&\quad\quad= (V_{pp}^\sigma - V_{pp}^\pi) \frac{\radio Y_{ji}}{\aC^2} \lsb \cos{\lrb\theta_i-\theta_j\rrb - 1}\rsb,\nonumber\\
 &&\rbra{z_i}\Ham\rket{z_j} = V_{pp}^\pi \cos{\lrb\theta_i -\theta_j\rrb} \nonumber\\
      &&\quad\quad-(V_{pp}^\sigma - V_{pp}^\pi) \frac{\radio^2}{\aC^2} \lsb \cos{\lrb\theta_i-\theta_j\rrb} - 1\rsb^2,
\end{eqnarray}
where $Y_{ji}=Y_j-Y_i$. This directly leads us to the final expression of the modified hopping parameters as given in Eq.~(\ref{eq:newt}).

\section{Approximations towards an effective Hamiltonian}\label{sec_appB}

The transfer integral $t_l $, which is a 2x2 matrix block as the spins are no longer degenerate, can be written as
\begin{equation}
 \begin{split}
  t_l &= \scriptstyle\lcb V_{pp}^\pi \lsb 1 - \frac{1}{2}\lrb \frac{d_l^\perp}{\radio}\rrb^2 \rsb
- (V_{pp}^\sigma - V_{pp}^\pi) \frac{\radio^2}{4\aC^2} \lrb \frac{d_l^\perp}{\radio}\rrb^4\rcb
  \begin{pmatrix}
   1 & 0\\
   0 & 1\\
  \end{pmatrix}\\
&+ 2\ii\delta\scriptstyle\lcb V_{pp}^\pi \frac{d_l^\perp}{\radio} + (V_{pp}^\sigma - V_{pp}^\pi) \frac{\radio^2}{2\aC^2} \lrb\frac{d_l^\perp}{\radio}\rrb^3\rcb
  \begin{pmatrix}
   1 & 0\\
   0 & -1\\
  \end{pmatrix}\\
&- 2\delta\scriptstyle(V_{pp}^\sigma - V_{pp}^\pi) \frac{\radio}{2\aC^2} \lsb 1 - \frac{\lrb d_l^\perp\rrb^2}{8 \radio^2}\rsb \lrb\frac{d_l^\perp}{\radio}\rrb^2 d_l^\parallel
  \begin{pmatrix}
   0 & \ee^{-\ii\tilde{\theta}}\\
   -\ee^{\ii\tilde{\theta}} & 0\\
  \end{pmatrix}\\
 \end{split}\label{eq:ts_matrix}
\end{equation}
where we have defined $\tilde{\theta} \equiv \frac{\theta_i + \theta_j}{2}$, and $d_l^\perp$ and $d_l^\parallel$ are the components of the $\vect{d}_l$ vector in the basis $\lcb \hat{x}_\perp, \hat{x}_\parallel\rcb$ of direction perpendicular and parallel to the tube axis, see Fig.~\ref{fig:coord}c.
The first row in the right-hand side of Eq.~(\ref{eq:ts_matrix}) contains only spin-independent elements, the second row groups spin-dependent terms whereas the last one is formed by a spin-flipping term. This last term depends on the local position of the interacting atoms in the tube through the angle $\tilde{\theta}$.

To obtain Eq.~(\ref{eq:ts_matrix}) from Eq.~(\ref{eq:newt}) we have taken the limit of large radii so that $\aC/\radio$ is a quantity small enough for the following substitutions to be reasonable:
\begin{itemize}
 \item  $\tilde{\theta}(\vect{R})\approx \theta\lrb\frac{\vect{R}_i + \vect{R}_j}{2}\rrb$. As seen in Fig.~\ref{fig:coord}b this is the case for a small enough $\lrb\theta_i-\theta_j\rrb$ angle.
 \item $\sin{\lrb\theta_i-\theta_j\rrb} \approx \frac{\vect{R}_{ij}\cdot \hat{x}_\perp}{\radio}=\frac{d_l^\perp}{\radio}$,\\
 $\cos{\lrb\theta_i-\theta_j\rrb} \approx  1 -\frac{\lrb d_l^\perp\rrb^2}{2 \radio^2}$,\\
\item $Y_{ij}=\vect{R}_{ij} \cdot \hat{x}_\parallel=d_l^\parallel$.
\end{itemize}

We will further expand the Hamiltonian around the Dirac points, $\Kone$ and $\Ktwo$, as we are interested in the physics taking place around the Fermi energy. The choice of
$\Kone=\frac{4\pi}{3\sqrt{3}\aC}(-1,0)$ and $\Ktwo=\frac{4\pi}{3\sqrt{3}\aC}(1,0)$ allows us to write the two independent Fermi points in a compact form as $\tau \Kone$ with $\tau=+1$ for $\Kone$ and $\tau=-1$ for $\Ktwo$.
For the expansion we substitute $\vect{k} \longrightarrow \tau \Kone + \vectGr{\kappa}$, $\vectGr{\kappa}$ being small enough to justify the expansion
$\ee^{\ii \vect{k} \cdot \vect{d}_l} = \ee^{\ii \tau\Kone \cdot \vect{d}_l} \lrb 1+\ii \vectGr{\kappa}\cdot \vect{d}_l\rrb$.

\vspace{1.2cm}

In this way the matrix block $\Ham_{AB}^\tau$ of the Hamiltonian reads
\begin{equation}
\begin{split}
&\sum_{l=1}^3 \ee^{\ii \vect{k} \cdot \vect{d}_l}\begin{pmatrix}t_{\uparrow\uparrow} & t_{\uparrow\downarrow}\\ t_{\downarrow\uparrow} & t_{\downarrow\downarrow} \end{pmatrix}
=  V_{pp}^\pi \frac{3\aC}{2} \ee^{-\ii \tau \eta} \Bigg[\lrb \tau \kappa_\perp - \ii \kappa_\parallel\rrb\\
&+ \frac{\aC\ee^{\ii \tau 3\eta}}{2^2 R^2}
- \frac{V_{pp}^\sigma - V_{pp}^\pi}{V_{pp}^\pi} \frac{\aC}{2^5 R^2} \lrb-4 \ee^{\ii \tau 3\eta} + \ee^{-\ii \tau 3\eta}\!\rrb \Bigg]
  \begin{pmatrix}
   1 & 0\\
   0 & 1\\
  \end{pmatrix}\\
&- V_{pp}^\pi \frac{3\aC}{2} \ee^{-\ii \tau \eta}\Bigg[ \delta \frac{2}{R}\tau
+ \delta \frac{V_{pp}^\sigma - V_{pp}^\pi}{V_{pp}^\pi}  \frac{3}{4 R} \tau \Bigg]
  \begin{pmatrix}
   1 &  0\\
   0 & -1\\
  \end{pmatrix}\\
&- V_{pp}^\pi \frac{3\aC}{2} \ee^{-\ii \tau \eta}\Bigg[ \delta \frac{V_{pp}^\sigma - V_{pp}^\pi}{V_{pp}^\pi} \frac{1}{4 R} \Bigg]
  \begin{pmatrix}
   0 & \ee^{-\ii\tilde{\theta}}\\
   -\ee^{\ii\tilde{\theta}} & 0\\
  \end{pmatrix}
\end{split}
\end{equation}
where we have kept terms up to the second order in the variables $\aC/\radio$, $\kappa$ and $\delta$ and neglected higher order terms.



\begin{thebibliography}{10}
\bibitem{notation}
We keep the conventional notation for the chirality of a \CNT, so that with our
  choice of lattice vectors a nanotube of chirality (n,m) has a chiral vector
  $\chiralvec=(n+m)\aone + m\atwo$.
\bibitem{limitsPeierls}
$\vect{A}$ is canceled out leaving us with the original Hamiltonian for zero
  magnetic field and no extra terms if assuming a spatially slow varying
  magnetic field and localized atomic functions.
\bibitem{length}
We impose the wave functions $\psi_\pm$ to vanish at the first row of missing
  atoms at the ends so that $L = \abso{\translvec} (N_{uc}+1/2)$, with
  $\translvec$ being the translational vector which has a length of
  $\sqrt{3}\aC$ for an armchair \CNTns, and $N_{uc}$ being the number of unit
  cells in the tube.
\end{thebibliography}
\end{document}